\documentclass[aps,showpacs,11pt]{revtex4}
\usepackage{dcolumn}
\usepackage{graphicx}
\usepackage{amsmath}
\usepackage{amsfonts}
\usepackage{amssymb}
\usepackage{psfrag}
\usepackage{wrapfig}
\usepackage{subfigure}
\usepackage{makeidx}
\usepackage{bm}
\usepackage{epsf}
\usepackage{hyperref}
\makeatletter

\newcommand{\Rmnum}[1]{\expandafter\@slowromancap\romannumeral #1@}
\makeatother

\begin{document}

\title{Critical behavior of Born-Infeld AdS black holes in the extended phase space thermodynamics}

\author{De-Cheng Zou}
\email{zoudecheng@sjtu.edu.cn}
\author{Shao-Jun Zhang}
\email{sjzhang84@sjtu.edu.cn}
\author{Bin Wang}
\email{wang$_$b@sjtu.edu.cn}

\affiliation{Department of Physics and Astronomy, Shanghai Jiao Tong University, Shanghai 200240, China}

\date{\today}

\begin{abstract}
\indent

We study the thermodynamics of $D$-dimensional Born-Infeld AdS black holes in the extended phase
space. We find that the usual small-large black hole phase transition, which exhibits analogy
with the Van de Waals liquid-gas system holds in all dimensions greater than three. However,
different from the four-dimensional case, in the system of higher dimensional Born-Infeld AdS
black holes there is no reentrant phase transition. For the three-dimensional Born-Infeld
AS black hole, there does not exist critical phenomena.

\end{abstract}


%\keywords{AdS/CFT correspondence, Phase transition, Born-Infeld black hole}

\maketitle

\section{Introduction}
\label{1s}

Black hole thermodynamics has been an intriguing subject of discussions for many years. It was
found that in the black hole spacetimes, we can define standard thermodynamic variables such as
temperature and entropy etc., and more strikingly we observe that black hole spacetimes contain
rich phase structures and admit critical phenomena which are in analogy with the
thermodynamic systems in nature.

Since the discovery of the AdS/CFT correspondence \cite{Maldacena:1997re,Gubser:1998bc,Witten:1998qj},
the study of thermodynamics in AdS black holes has become more attractive nowadays. In view of
the AdS/CFT correspondence, the bulk AdS black hole spacetimes admit a gauge duality description
by thermal conformal field theory living on the AdS boundary. The first investigation of the
thermodynamic properties in AdS black holes was reported in \cite{Hawking:1982dh}, where it was
demonstrated that a certain phase transition in the phase space of the Schwarzschild AdS black
hole exists. This phase transition can be interpreted as a confinement/deconfinement phase
transition in the dual quark gluon plasma in the language of AdS/CFT correspondence \cite{Witten:1998zw}.
More interesting discoveries were obtained for the charged AdS black holes \cite{Chamblin:1999tk, Chamblin:1999hg},
where it was found that a first order small black hole and big black hole phase transition is allowed which is
superficially analogous to a liquid-gas phase transition of the Van der Waals fluid. This
superficial reminiscence was also observed in other AdS backgrounds
\cite{Banerjee:2011au,Cao:2010ft, Myung:2008eb,Banerjee:2011cz,
Fernando:2006gh, Poshteh:2013pba,Mo:2013sxa,Dey:2007vt,Niu:2011tb,Tsai:2011gv,Lu:2010au,Lala:2012jp,Banerjee:2012zm}.

Recently the study of thermodynamics in AdS black holes has been generalized to the extended phase
space, where the cosmological constant is identified with thermodynamic pressure and its variations are included
in the first law of black hole thermodynamics \cite{Dolan:2011xt,Dolan:2010ha,Dolan:2011jm,Dolan:2013ft,Spallucci:2013osa}.
Including the variable cosmological constant in the first law was motivated by the geometric
derivations of the Smarr formula \cite{Kastor:2009wy}.  In so doing the AdS black hole mass
is identified with enthalpy and there exists a natural conjugate thermodynamic volume to the
cosmological constant. In the extended phase space with cosmological constant and volume as
thermodynamic variables, it was interestingly found that the system admits a more direct and
precise coincidence between the first order small-large black hole phase transition and the
liquid-gas change of phase occurring in fluids \cite{Kubiznak:2012wp}. Considering the extended
phase space, and hence treating the cosmological constant as a dynamical quantity, is a very
interesting theoretical idea in disclosing possible phase transitions in AdS black holes \cite{Gunasekaran:2012dq}.
More discussions in this direction can be found as well in
\cite{Hendi:2012um, Belhaj:2013ioa, Belhaj:2012bg, Chen:2013ce, Cai:2013qga,Zhao:2013oza,Belhaj:2013cva,  Xu:2013zea, Dutta:2013dca}.

Generalizing the discussion of the extended phase space thermodynamics to higher dimensional singly
spinning AdS black holes, it was uncovered that besides the usual small-large black hole phase
transitions, there exists a new phenomenon of reentrant phase transitions for all $d\geq 6$
dimensions, in which a monotonic variation of the temperature yields two phase transitions from
large to small and back to large black holes \cite{Altamirano:2013ane}. This new reentrant
phase transition was also observed in the multi-spinning $D=6$ Kerr-anti-de Sitter black hole with fixed
angular momenta $J_1$ and $J_2$ \cite{Altamirano:2013uqa}. It was argued in \cite{Altamirano:2013ane, Altamirano:2013uqa}
that the new phase structure in the thermodynamics of rotating black holes resembles to binary fluids
seen in superfluidity and superconductivity.
In addition to including the rotation, especially in higher dimensions where one has more rotation
parameters, which can make the phase transitions more interesting, the impact of possible
non-linear electrodynamics extensions of RN-AdS black hole on the extended phase space
has also been examined. But this study was restricted in the four-dimensional
Born-Infeld AdS black holes \cite{Gunasekaran:2012dq}. To be consistent with
the corresponding Smarr relation \cite{Kubiznak:2012wp},
it was argued that further extension of the phase space is needed by including the
variation of the maximal field strength $b$ in the first law of thermodynamics. The
thermodynamic quantity conjugate to the Born-Infeld parameter $b$ is called Born-Infeld vacuum polarization.
In four-dimensional Born-Infeld AdS black holes \cite{Gunasekaran:2012dq}, it was found that the
impact of the nonlinearity can bring the new phenomenon of reentrant phase transition which
was observed in rotating AdS holes \cite{Altamirano:2013ane, Altamirano:2013uqa}.

It is interesting to generalize the discussion of the impact of non-linear electrodynamics on the
extended phase space thermodynamics to other spacetime dimensions and examine whether the
phase structures observed in four-dimensional Born-Infeld AdS black holes are general. This is
the motivation of the present paper. We will show that the four-dimensional Born-Infeld AdS black
hole is very special. When the spacetime dimension is greater than four, there only exists
usual small-large black hole phase transition but no new reentrant phase transition. In addition,
in three dimensional Born-Infeld AdS black holes, we cannot find any critical phenomena. Without
loss of generality, we will discuss the $D$-dimensional Born-Infeld AdS black holes with
different topologies, such as Ricci flat,
hyperbolic or spherical hyper-surfaces.

This paper is organized as follows. In Sec.~\ref{2s}, we will discuss the solutions of
$D$-dimensional Born-Infeld AdS black holes. In Sec.~\ref{3s}, we will investigate the phase
transitions of $D$-dimensional Born-Infeld AdS black holes. We will summarize our results in Sec.~\ref{4s}.

\section{Solutions of $D$-dimensional Born-Infeld AdS black holes}
\label{2s}

The action of Einstein gravity in the presence of Born-Infeld field reads  \cite{Banerjee:2012zm}
\begin{eqnarray}
\mathcal{I} =\frac{1}{16\pi}\int d^{D}x \sqrt{-g}\left[-2\Lambda+R+{\cal L(F)}\right],\label{eq:1a}
\end{eqnarray}
where $\Lambda=-\frac{(D-1)(D-2)}{2l^2}$ is the negative cosmological constant, $b$ is the
Born-Infeld parameter and ${\cal L(F)}$ is given by
\begin{eqnarray}
{\cal L(F)}=4b^2(1-\sqrt{1+\frac{F^{\mu\nu}F_{\mu\nu}}{2b^2}}). \label{eq:2a}
\end{eqnarray}
In the limit $b\rightarrow \infty$, it reduces to the standard Maxwell field ${\cal
L(F)}=-F^{\mu\nu}F_{\mu\nu}+\mathcal {O}(F^4)$. If we take $b=0$, ${\cal L(F)}$ disappears.

We assume the non-rotating metric in the form
\begin{eqnarray}
ds^2=-f(r)dt^2+\frac{1}{f(r)}dr^2+r^2h_{ij}dx^idx^j,\label{eq:3a}
\end{eqnarray}
where the coordinates are labeled as $x^{\mu}=(t,r, x^i)$, $(i=1,\cdots ,D-2)$. $h_{ij}$ describes
the $(D-2)$-dimensional hypersurface with constant scalar curvature $(D-2)(D-3)k$. The
constant $k$ characterizes the topology which can be $k=0$ (flat), $k=-1$ (negative curvature) and
$k=1$ (positive curvature). The solution is given by \cite{Cai:2004eh, Dey:2004yt}
\begin{eqnarray}
f(r)&=&k+\frac{r^2}{l^2}-\frac{m}{r^{D-3}}+\frac{4b^2r^2}{(D-1)(D-2)}\left(1
-\sqrt{1+\frac{(D-2)(D-3)q^2}{2b^2r^{2D-4}}}\right)\nonumber\\
&&+\frac{2(D-2)q^2}{(D-1)r^{2n-4}}{_2}F_1[\frac{D-3}{2D-4},\frac{1}{2},\frac{3D-7}{2D-4},
-\frac{(D-2)(D-3)q^2}{2b^2 r^{2D-4}}], \label{eq:4a}\\
F&=&dA, \quad A=-\sqrt{\frac{D-2}{2(D-3)}}\frac{q}{r^{D-3}}{_2}F_1[\frac{D-3}{2D-4},\frac{1}{2},
\frac{3D-7}{2D-4},-\frac{(D-2)(D-3)q^2}{2b^2 r^{2D-4}}]dt, \label{eq:5a}
\end{eqnarray}
where $m$ and $q$ are related to the mass $M$ and charge $Q$ of black holes as
\begin{eqnarray}
Q=\frac{q\Sigma_k }{4\pi}\sqrt{\frac{(D-2)(D-3)}{2}}, \quad M=\frac{(D-2)\Sigma_k}{16\pi}m.\label{eq:6a}
\end{eqnarray}
Here $\Sigma_k$ represents the volume of constant curvature hypersurface described by
$h_{ij}dx^idx^j$. The electromagnetic potential difference ($\Phi$) between the horizon and infinity is
\begin{eqnarray}
\Phi=\frac{4\pi Q}{(D-3)\Sigma_k r_+^{D-3}}{_2}F_1[\frac{D-3}{2D-4},\frac{1}{2},\frac{3D-7}{2D-4},
-\frac{16\pi^2 Q^2}{b^2\Sigma_k^2 r_+^{2D-4}}].\label{eq:7a}
\end{eqnarray}

In terms of the horizon radius $r_+$, the ADM mass $M$ of the Born-Infeld AdS black holes is obtained
\begin{eqnarray}
M&=&\frac{(D-2)\Sigma_kr_+^{D-3}}{16\pi}\left[k+\frac{r_+^2}{l^2}
+\frac{4b^2r_+^2}{(D-1)(D-2)}\left(1-\sqrt{1+\frac{16\pi^2 Q^2}{b^2\Sigma_k^2 r_+^{2D-4}}}\right)\right.\nonumber\\
&&\left.+\frac{64\pi^2 Q^2}{(D-1)(D-3)\Sigma_k^2r_+^{2D-6}}{_2}F_1[\frac{D-3}{2D-4},\frac{1}{2},\frac{3D-7}{2D-4},
-\frac{16\pi^2 Q^2}{b^2\Sigma_k^2 r_+^{2D-4}}]\right].\label{eq:8a}
\end{eqnarray}
The Hawking temperature and entropy of the Born-Infeld AdS black holes are
\begin{eqnarray}
T&=&\frac{1}{4\pi}\left[\frac{(D-1)r_+}{l^2}+\frac{(D-3)k}{r_+}
+\frac{4b^2r_+}{D-2}\left(1-\sqrt{1+\frac{16\pi^2 Q^2}{b^2\Sigma_k^2 r_+^{2D-4}}}\right)\right],\label{eq:9a}\\
S&=&\frac{\Sigma_k}{4} r_+^{D-2}. \label{eq:10a}
\end{eqnarray}

%%%%%%%%%%%%start from here%%%%%%%%%%%%%%%
From the temperature expression, we can derive the condition to accommodate the extremal black
holes with $T=0$. For $D=4$, the vanishing of $T$ requires
\begin{eqnarray}
k+\left(\frac{3}{l^2}+2b^2\right)r_{ext}^2-2b\sqrt{b^2 r_{ext}^{4}+Q^2}=0,\label{eq:11a}
\end{eqnarray}
which leads two possible solutions of the horizon radius for the extremal black holes
\begin{eqnarray}
r_{ext1,2}^2=\frac{l^2(1+\frac{3}{2b^2l^2})}{6(1+\frac{3}{4b^2l^2})}\left[-k\pm\sqrt{k^2+\frac{12(1+\frac{3}{4b^2l^2})}{b^2l^2(1
+\frac{3}{2b^2l^2})^2}(b^2Q^2-\frac{k^2}{4})}\right].\label{eq:12a}
\end{eqnarray}
For the spherical case with $k=1$, one can see that the solution with $``-"$ sign in front of
the square root is always unphysically negative. For the solution with plus sign in front of the
square root, when $bQ\geq1/2$ it is real; when $0<bQ<1/2$ this solution is imaginary. This shows
that for $k=1$, only when $bQ\geq1/2$, it is possible to have two horizon enveloping the central singularity
and these two horizon degenerate for the extreme case with $T=0$ for the black hole.
When $0<bQ<1/2$, the black hole can never become extreme. For the hyperbolic
topology $k=-1$, although the solution of $r_{ext}^2$ with $``-"$ sign in front of the
square root is unphysically negative, $r_{ext}^2$ with $``+"$ sign in front of the square root is
always real and positive no matter what value of $bQ$ we take, since
$1+\frac{12(1+\frac{3}{4b^2l^2})}{b^2l^2(1+\frac{3}{2b^2l^2})^2}(b^2Q^2-\frac{1}{4})
=\frac{4b^2\left(9Q^2+b^2l^2(l^2+12Q^2)\right)}{(3+2b^2l^2)^2}>0$.
Thus when $k=-1$, the Born-Infeld AdS black hole exists for all values of $bQ$ and this
black hole can become extreme to allow $T=0$.
For the black hole with flat topology $k=0$, there exists one real positive root
$r_{ext}^2=\frac{2bQl^2}{\left[3(3+4b^2l^2)\right]^{1/2}}$ to ensure $T=0$, which shows that for the flat case,
no matter what value of $bQ$ we choose, Born-Infeld AdS black hole can become extreme.
The behaviors of the black hole temperature $T$ versus the mass $m$ are plotted for fixed $Q=1$,
$l=1$ and different values of $b$ in four-dimensions in Fig.~1, where it shows that
there does not exist any extremal spherical Born-Infeld AdS black holes with $T=0$ when $0<bQ<1/2$.

%%%%%%%%%%%%%%%%%%%%%%%%%%%%%%%%%%%%%%%%%%%%%%%%%%%%%%%%%%%%%%%%%%%%%%%%%%%
\begin{figure}[h]
\centering
\subfigure[$k=1$]{
\label{fig:subfig:a} %% label for first subfigure
\includegraphics[width=0.3\textwidth]{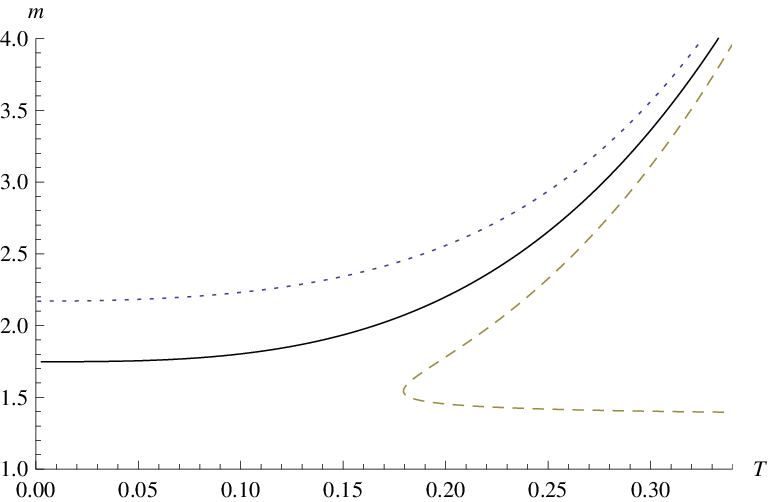}}%
\hfill%
\subfigure[$k=-1$]{
\label{fig:subfig:b} %% label for first subfigure
\includegraphics[width=0.3\textwidth]{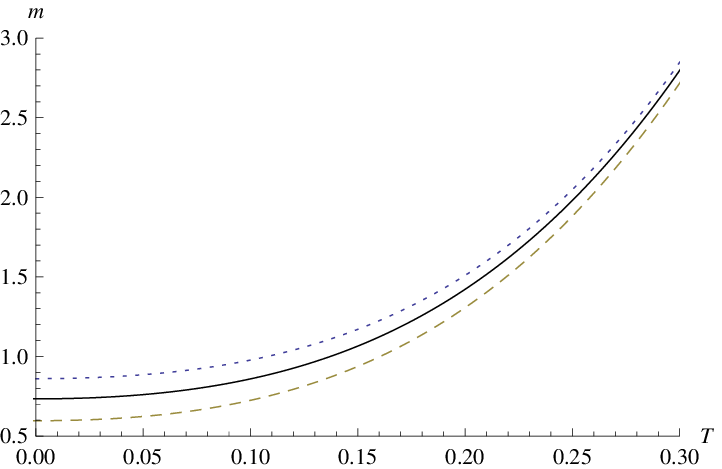}}%
\hfill%
\subfigure[$k=0$]{
\label{fig:subfig:c} %% label for second subfigure
\includegraphics[width=0.3\textwidth]{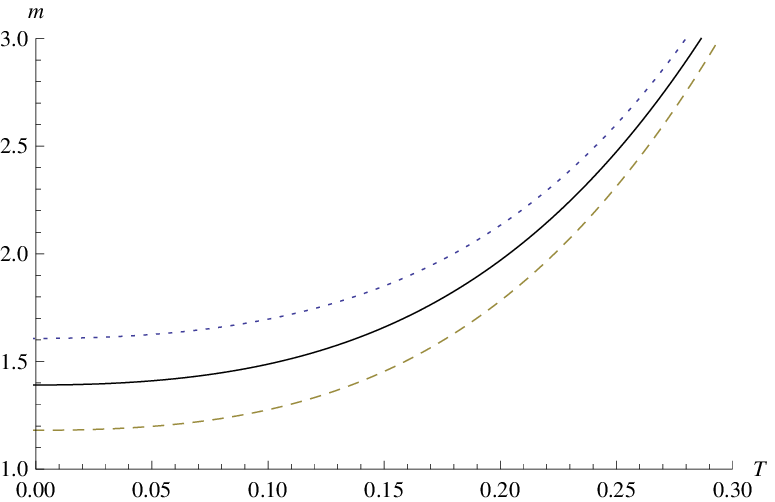}}
\caption{The behaviors of temperatures of black
holes with different topologies versus mass
parameters $m$. We fix $Q=1$, $l=1$, $D=4$. The
lines from above to below correspond to $b=1$,
$0.5$ and $0.3$, respectively. }
\label{fig:subfig} %% label for entire figure
\end{figure}

When the spacetime dimension is higher than four, to accommodate $T=0$, we require
\begin{eqnarray}
\frac{(D-1)r_{ext}^2}{l^2}+(D-3)k
+\frac{4b^2r_{ext}^2}{D-2}\left(1-\sqrt{1+\frac{16\pi^2 Q^2}{b^2\Sigma_k^2 r_{ext}^{2D-4}}}\right)=0.\label{eq:13a}
\end{eqnarray}
For the black hole with flat topology, there always exists one real positive root
$r_{ext}=\left[\frac{256l^4\pi^2b^2Q^2}{(D-1)(D-2)\left((D-1)(D-2)+8b^2l^2\right)\Sigma_k}\right]^{2D-4}$
and positive mass parameter $m$ of extremal higher dimensional black hole to ensure $T=0$,
which is same as the case in four dimensional spacetimes.
However, it is difficult to obtain the analytic solution of this equation
for the cases of $k=1$ and $k=-1$. We can only present numerical results of the mass parameter $m$ when $T=0$
for $D=5$ in the TABLE.~\Rmnum{1}. Note that for spherical topology, no matter what value of $bQ$,
we can always have extreme black holes with $T=0$. This implies that the five dimensional
case is different from the four-dimensional case, there is no limit for the value of $bQ$ to allow
the existence of the Born-Infeld AdS black hole even for spherical topology.
For the five dimensional hyperbolic Born-Infeld AdS black hole with $k=-1$, however,
$T=0$ can only be achieved for negative $m$. Since $m$ is associated with the mass $M$, according to Eq. (6),
this implies a negative mass of the spacetime, which is not physical. Thus we restrict our
consideration to $m>0$ so that the black hole can never become extreme.
In Fig.~2 we show the temperatures $T$ of black holes with $m$ for different topologies for $D=5$.

\begin{table}[h]
\begin{tabular}{|c||c||c||c||c||c||c||c||c||c||c|}
  \hline
   parameters & \multicolumn{5}{|c|}{$k=1$} & \multicolumn{5}{|c|}{$k=-1$}\\ \hline
  $b $ & 0.01 & 0.3 & 0.5& 1 & 10 & 0.01 & 0.3 & 0.5& 1 & 10 \\ \hline
  $r_{ext}$ & 0.0042 &0.1234 & 0.1960& 0.3296 & 0.5360 & 0.7092 & 0.7543 & 0.7723 &0.7937 & 0.8101 \\ \hline
  $m$ &0.0357 & 0.3282 & 0.4406&  0.6124 & 0.8353 &-0.2204 & -0.0840& -0.0581 & -0.0343 & -0.0200 \\ \hline
\end{tabular}
\caption{The numerical results of these parameters $r_{ext}$ and $m$ of five dimensional Born-Infeld AdS black holes
with different topologies. We take $Q=1$ and $l=1$. }
\end{table}

%%%%%%%%%%%%%%%%%%%%%%%%%%%%%%%%%%%%%%%%%%%%%%%%%%%%%%%%%%%%%%%%%%%%%%%%%%%
\begin{figure}[h]
\centering
\subfigure[$k=1$]{
\label{fig:subfig:a} %% label for first subfigure
\includegraphics[width=0.3\textwidth]{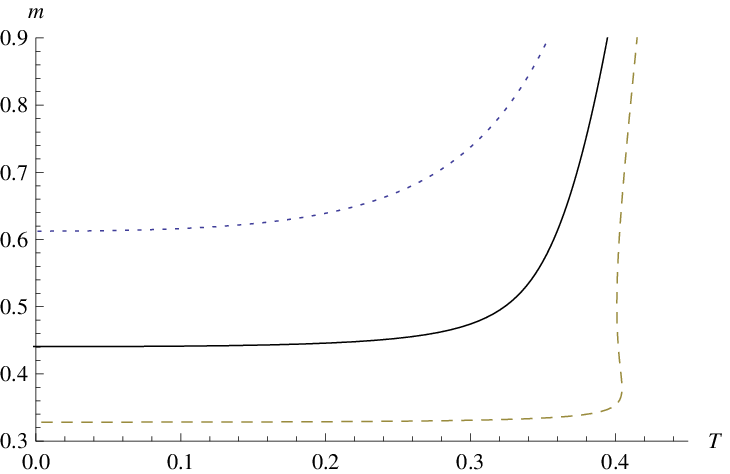}}%
\hfill%
\subfigure[$k=-1$]{
\label{fig:subfig:b} %% label for first subfigure
\includegraphics[width=0.3\textwidth]{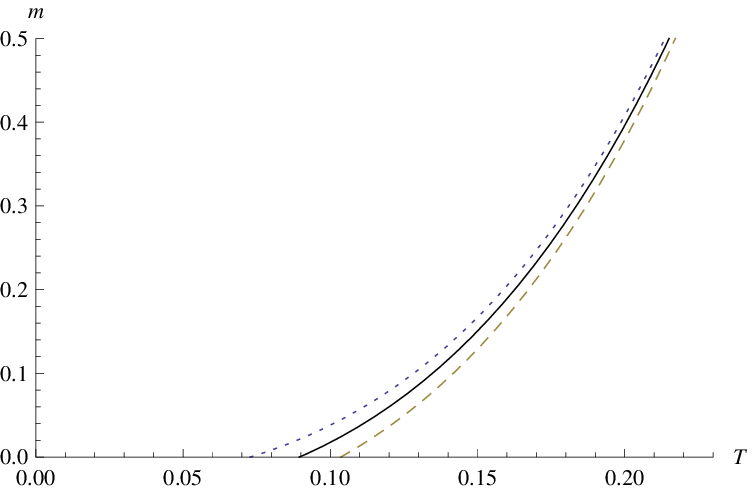}}%
\hfill%
\subfigure[$k=0$]{
\label{fig:subfig:c} %% label for second subfigure
\includegraphics[width=0.3\textwidth]{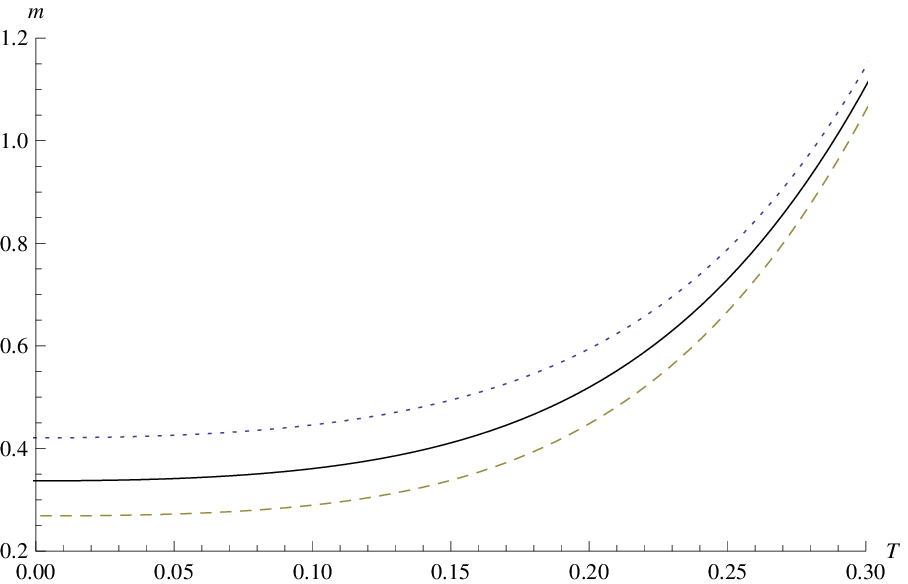}}
\caption{ The behaviors of mass parameters $m$ of black holes with different topologies versus temperatures $T$.
We fix $Q=1$, $l=1$, $D=5$. The lines from up to below correspond to $b=1$, $0.5$ and $0.3$, respectively. }
\end{figure}

To see more closely of the behaviors of the Born-Infeld AdS black hole, we examine the function $f(r)$.
The black hole possesses a singularity at $r=0$, which can be a naked singularity or enveloped by one, or
two horizons, depending on the values of the parameters. The expansion of the function $f(r)$ around $r=0$
takes the form \cite{Dey:2004yt}
\begin{eqnarray}
f(r)=k-\frac{m-A}{r^{D-3}}-\left(\frac{2Cb}{D-1}-B\right)\sqrt{\frac{2}{(D-2)(D-3)}}\frac{4\pi Q}{\Sigma_k r^{D-4}}
+\mathcal{O}(r),\label{eq:14a}
\end{eqnarray}
where
\begin{eqnarray}
A&=&\frac{64\pi^2Q^2}{(D-1)(D-3)\sqrt{\pi}\Sigma_k^2}\left(\frac{b^2\Sigma_k^2}{16\pi^2Q^2}\right)^{\frac{D
-3}{2D-4}}\Gamma[\frac{3D-7}{2D-4}]\Gamma[\frac{1}{2D-4}],\nonumber\\
B&=&\frac{4b}{(D-1)C}\frac{\Gamma[\frac{3D-7}{2D-4}]\Gamma[\frac{-1}{2D-4}]}{\Gamma[\frac{D
-3}{2D-4}]\Gamma[\frac{2D-5}{2D-4}]},\quad C=\sqrt{\frac{2D-6}{D-2}}.\label{eq:15a}
\end{eqnarray}

For $D=4$, Eq.~(\ref{eq:14a}) can be simplified as \cite{Myung:2008kd, Fernando:2006gh}
\begin{eqnarray}
f(r)=k-\frac{m-A}{r}-2bQ+\mathcal{O}(r), \quad A=\frac{1}{3}\sqrt{\frac{b}{\pi}}Q^{3/2}\Gamma(\frac{1}{4})^2.\label{eq:16a}
\end{eqnarray}
For the spherical case with $k=1$, the behavior of the solution was discussed in details in \cite{Gunasekaran:2012dq}.
If we take $m>A$, the function $f(r)=1-\frac{m-A}{r}-2bQ+\mathcal{O}(r)$ approaches $-\infty$ near the origin $r=0$ and
approaches  $+\infty$ when $r\rightarrow+\infty$, which is independent of the value of
$bQ$. The behavior of the Born-Infeld AdS resembles the ``Schwarzschild-like" (S) type with a spacelike singularity and
one horizon enveloping it. If we take $m<A$, the function $f(r)=1-\frac{m-A}{r}-2bQ+\mathcal{O}(r)$
approaches $+\infty$ near the origin $r=0$. In this case, only when $bQ>1/2$ we can find the solutions of $r$ to satisfy
$f(r)=1-\frac{m-A}{r}-2bQ+\mathcal{O}(r)=0$, which describe the horizons of  the black hole.
The behavior of $f(r)$ likes the ``Reissner-N$\ddot{o}$rdstrom" (RN) type, which can have two horizons
if $m_{ext}< m<A$ and these two horizons degenerate when $m_{ext}=A$, while the RN type describes a
naked singularity when $ m<m_{ext}<A$. When $bQ<1/2$, the $f(r)$ only possesses the S-type
black holes when $m\geq A$. This is consistent with the analysis of the temperature.  These
properties are shown in Fig.3a. It is interesting to note that $m=A$ describes the
`marginal' case, where the function $f(r)$ Eq.~(\ref{eq:16a}) reduces to
$f_A=1-2bQ+\mathcal{O}(r)$. Obviously the function $f_A$ possesses one horizon for
$bQ>1/2$; but when $bQ<1/2$, $f_A$ is positive and describes a naked singularity, see Fig.~4a.

For the topologies with $k=0$ and $-1$, when $m>A$, the function
$f(r)=k-\frac{m-A}{r}-2bQ+\mathcal{O}(r)$ tends to $-\infty$ near $r=0$ and approaches $+\infty$
when $r\rightarrow +\infty$, so that the Born-Infeld AdS black hole possesses the property
of the S-type black hole. When $m<A$, $f(r)$ attains $+\infty$ near $r=0$ and $+\infty$
as $r\rightarrow +\infty$. No matter what positive values of $bQ$ we take, the black hole
can always have the RN-type when $m<A$ for $k=0$ and $-1$. This is different from
the spherical case and agrees with the observations we got in the temperature
expression. The result is shown in Figs.~3(b)(c). As to the `marginal' case $m=A$, the function
$f(r)$ can be written as $f_A=k-2bQ+\mathcal{O}(r)$, which is always
negative at $r=0$ and takes $+\infty$ as $r\rightarrow +\infty$. This shows that for the
marginal case the black hole always possesses a horizon for positive values of $bQ$, see Figs.~4(b)(c).
Finally, these different types for the black hole solution $f(r)$ Eq.~(\ref{eq:4a}) in four
dimensional spacetimes are shown in Fig.~5.

%%%%%%%%%%%%%%%%%%%%%%%%%%%%%%%%%%%%%%%%%%%%%%%%%%%%%%%%%%%%%%%%%%%%%%%%%%%
\begin{figure}[htb]
\centering
\subfigure[$k=1$]{
\label{fig:subfig:a} %% label for first subfigure
\includegraphics[width=0.3\textwidth]{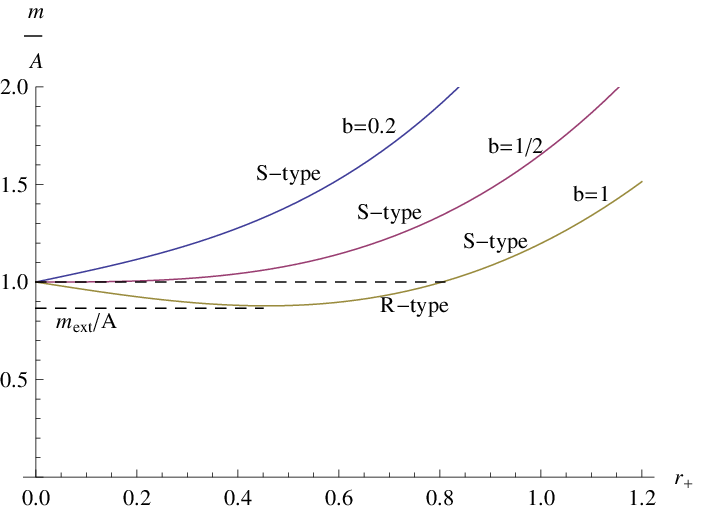}}%
\hfill%
\subfigure[$k=-1$]{
\label{fig:subfig:b} %% label for first subfigure
\includegraphics[width=0.3\textwidth]{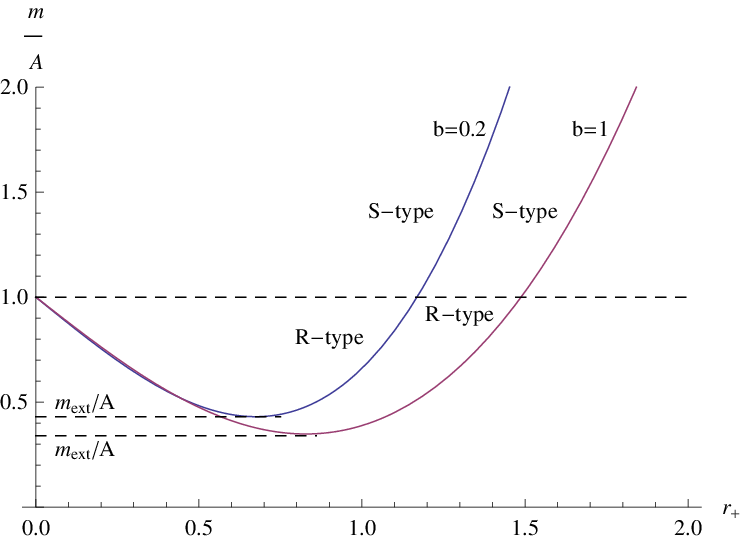}}%
\hfill%
\subfigure[$k=0$]{
\label{fig:subfig:b} %% label for second subfigure
\includegraphics[width=0.3\textwidth]{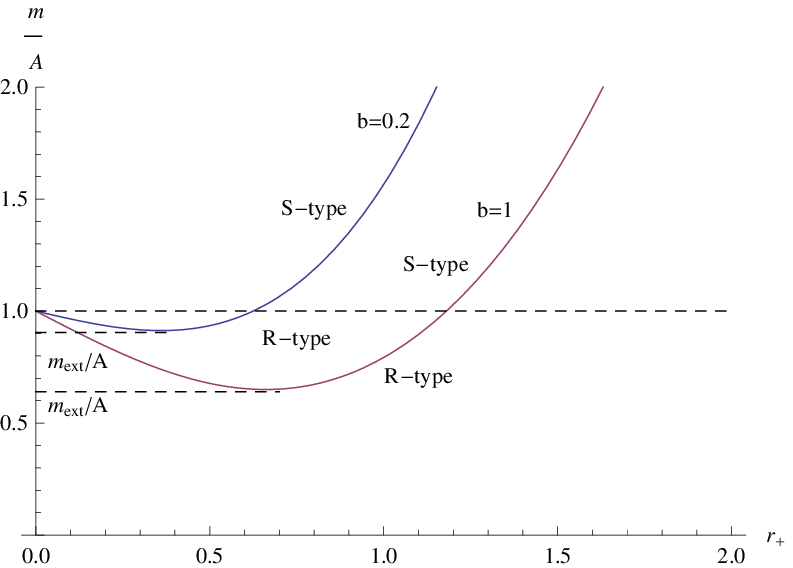}}
\caption{ Solutions of black hole horizons in
four-dimensions as functions of $m/A$. We take
$Q=1$, $l=1$. }
\label{fig:subfig} %% label for entire figure
\end{figure}

%%%%%%%%%%%%%%%%%%%%%%%%%%%%%%%%%%%%%%%%%%%%%%%%%%%%%%%%%%%%%%%%%%%%%%%%%%%
\begin{figure}[htb]
\centering
\subfigure[$k=1$]{\label{fig:a} %% label for second subfigure
\includegraphics[width=0.3\textwidth]{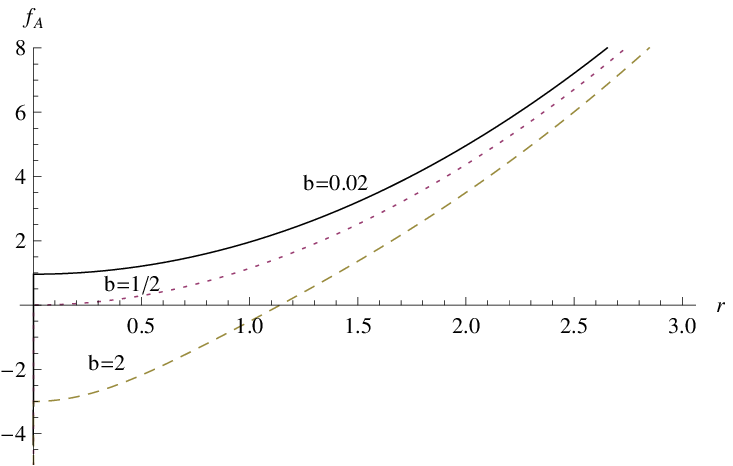}}%
\hfill%
\subfigure[$k=-1$]{\label{fig:b} %% label for second subfigure
\includegraphics[width=0.3\textwidth]{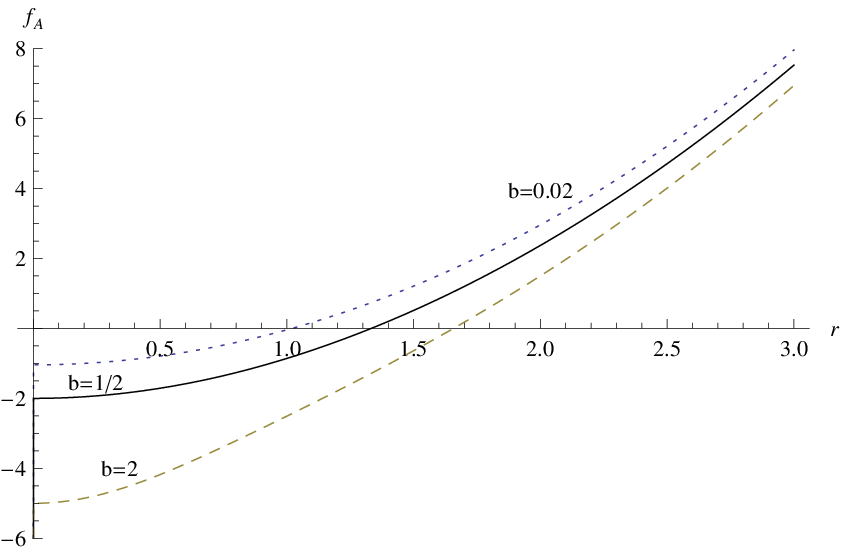}}%
\hfill%
\subfigure[$k=0$]{\label{fig:c} %% label for second subfigure
\includegraphics[width=0.3\textwidth]{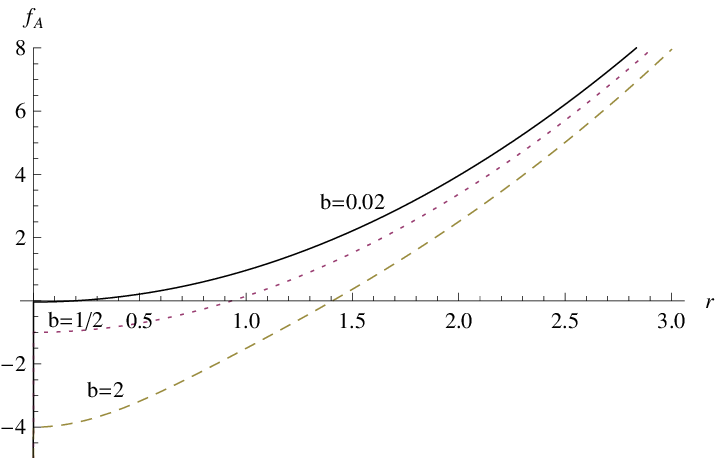}}%
\caption{ Marginal case in four-dimensions. We
fix $Q=1$, $l=1$.}
\label{fig:subfig} %% label for entire figure
\end{figure}

%%%%%%%%%%%%%%%%%%%%%%%%%%%%%%%%%%%%%%%%%%%%%%%%%%%%%%%%%%%%%%%%%%%%%%%%%%%
\begin{figure}[htb]
\centering
\subfigure[$k=1$]{\label{fig:a} %% label for second subfigure
\includegraphics[width=0.3\textwidth]{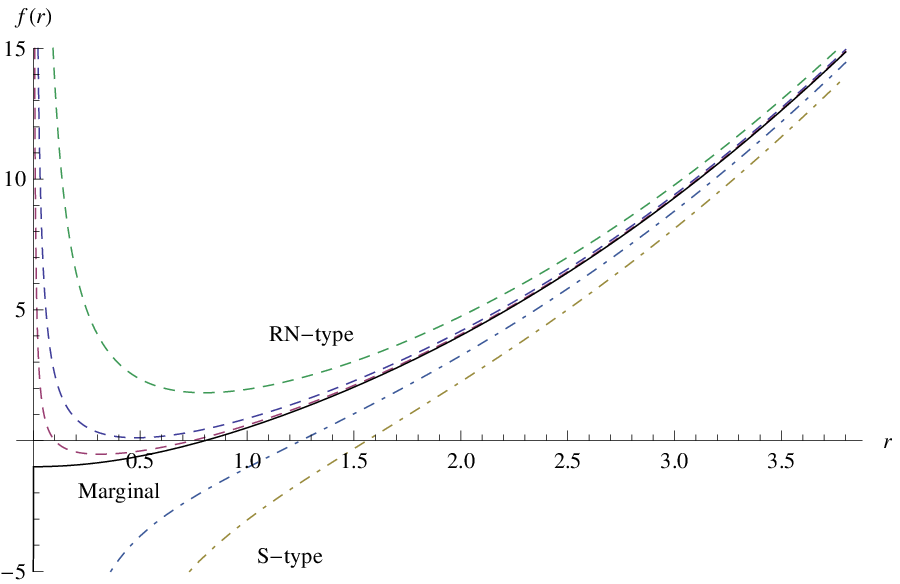}}%
\hfill%
\subfigure[$k=-1$]{\label{fig:b} %% label for second subfigure
\includegraphics[width=0.3\textwidth]{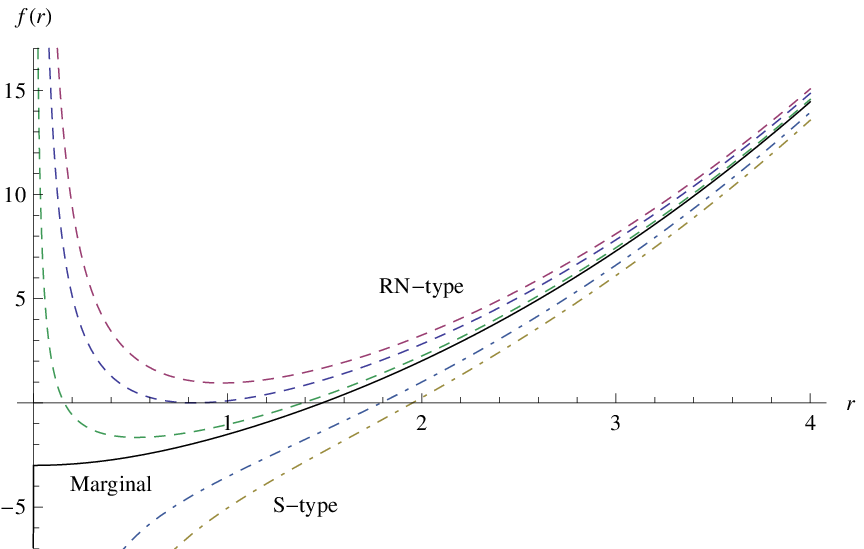}}%
\hfill%
\subfigure[$k=0$]{\label{fig:c} %% label for second subfigure
\includegraphics[width=0.3\textwidth]{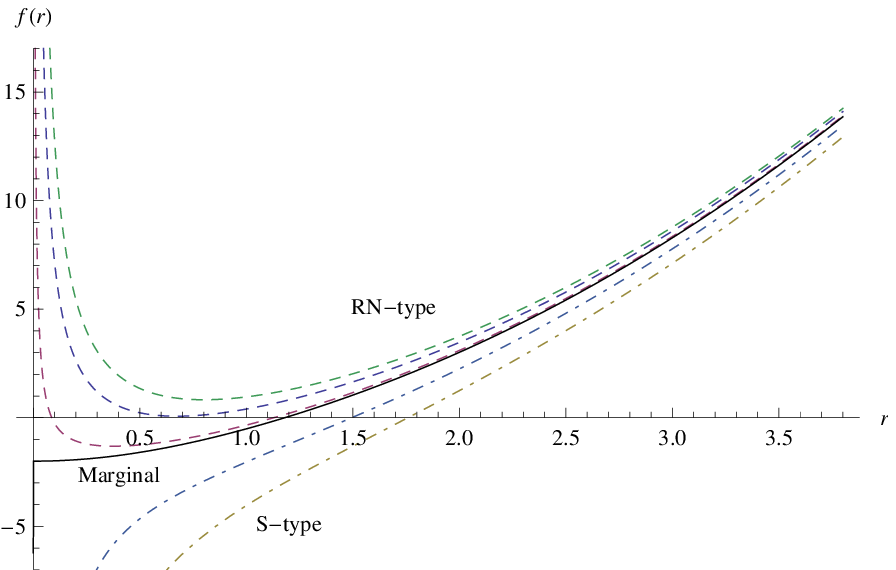}}%
\caption{Types of the four-dimensional Born-Infeld
AdS black holes with different topologies. The
marginal type of black hole corresponds to $m=A$,
the S-type black hole corresponds to $m>A$ and
the RN-type of black hole is for $m<A$.
For the RN-type black hole, the three dashed lines from top to below
correspond to the naked singularity taking $m<m_{ext}<A$,
extremal BH $m=m_{ext}<A$ and two horizon solutions
$m_{ext}<m<A$ respectively. In
plotting the figure, we have used $Q = 1, b =1$.}
\label{fig:subfig} %% label for entire figure
\end{figure}

Now we turn our discussion to higher dimensional Born-Infeld AdS black holes. Taking $D=5$,
Eq.~(\ref{eq:14a}) becomes
\begin{eqnarray}
f(r)=k-\frac{m-A'}{r^2}-\frac{b'Q}{r}+\mathcal{O}(r), \quad b'=\frac{58\sqrt{2}b}{25\pi},
\quad A'=\frac{(2b^2Q^4)^{1/3}}{\pi^{11/6}}\Gamma(\frac{4}{3})\Gamma(\frac{1}{6}). \label{eq:17a}
\end{eqnarray}
For all topological cases when $m>A'$, we see that the function $f(r)$ tends to $-\infty$ near
$r=0$ and $+\infty$ when $r\rightarrow +\infty$. Therefore the behavior of $f(r)$ describes the
Schwarzschild type black hole. If we take $m<A'$, the function $f(r)$
approaches $+\infty$ when $r\rightarrow 0$ and $r\rightarrow +\infty$. This asymptotical
property keeps for all values of $k$. For small $r$, $k$ is negligible compared with $b'Q/r$,
thus for all positive values of $b'Q$ the black hole can possess the RN type which may have two
horizons when $m_{ext}< m<A'$ and only one degenerated horizon when $m_{ext}=A'$. For the
spherical case, when the spacetime dimension is higher than four, the restriction of $bQ$ to
accommodate the RN-type Born-Infeld AdS black hole is washed out. This is in agreement with the
analysis of the temperature expression. When $k=-1$, there
does not exist extreme black hole and when the mass parameter $m$ within the range $0<m<A'$, it
describes the RN-type black hole. The numerical picture is shown in Fig.~6.

When $m=A'$, Eq.~(\ref{eq:17a}) reduces to
\begin{eqnarray}
f_{A'}=f(m=A')=k-\frac{b'Q}{r}+\mathcal{O}(r).\label{eq:18a}
\end{eqnarray}
Different from the case of $D=4$, the term $\frac{b'Q}{r}$ is very large for small $r$, so
that $f_{A'}\rightarrow -\infty$ near $r=0$ and $f_{A'}\rightarrow+\infty$ as $r\rightarrow
+\infty$, provided that $b'Q>0$. The ``marginal" case $m=A'$ is characterized by the existence of
a spacelike singularity enveloped by a horizon for all different topologies, see Fig.~7.
Then these different types for the black hole solution $f(r)$ Eq.~(\ref{eq:4a}) in five
dimensional spacetimes are shown in Fig.~8, where the ``marginal" case $(m=A')$ also
belongs to the S-type, which is drawn with black line. For the RN-type black hole $(m<A')$ with $k=1$ and 0,
the three dashed lines from top to below correspond to the naked singularity taking $m<m_{ext}<A'$,
extremal BH $m=m_{ext}<A'$ and two horizon solutions $m_{ext}<m<A'$ respectively.
It is interesting to note that since there does not exist extremal black
hole for five dimensional hyperbolic spacetime,
the Born-Infeld AdS black hole always possesses two horizons in the form of RN-type.
When $D>5$, $\left(\frac{2Cb}{D-1}-B\right)\sqrt{\frac{2}{(D-2)(D-3)}}\frac{4\pi Q}{\Sigma_k}$
always maintains positive for $k=0,
1$ and $-1$, one can see that the behavior of $f(r)$ is similar to the case of $D=5$.

%%%%%%%%%%%%%%%%%%%%%%%%%%%%%%%%%%%%%%%%%%%%%%%%%%%%%%%%%%%%%%%%%%%%%%%%%%%
\begin{figure}[h]
\centering
\subfigure[$k=1$]{
\label{fig:subfig:a} %% label for first subfigure
\includegraphics[width=0.3\textwidth]{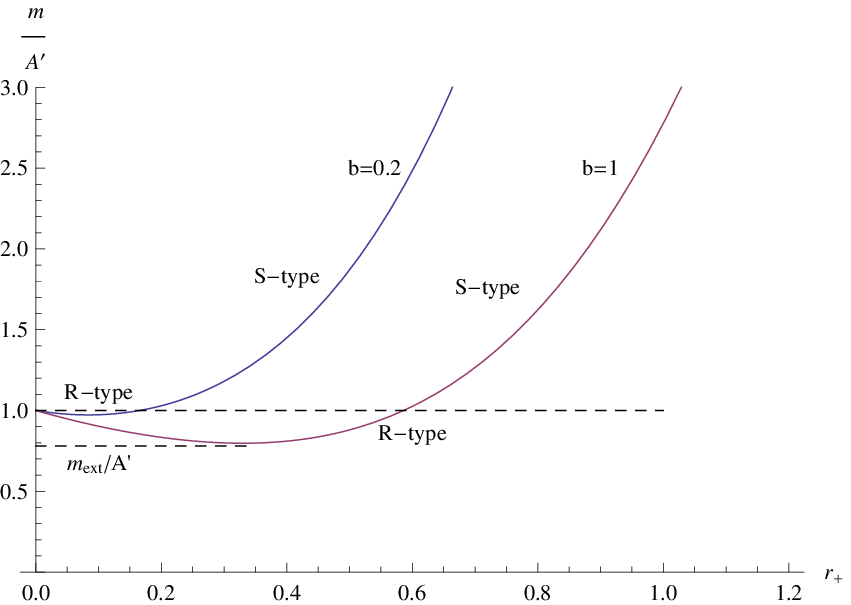}}%
\hfill%
\subfigure[$k=-1$]{
\label{fig:subfig:b} %% label for first subfigure
\includegraphics[width=0.3\textwidth]{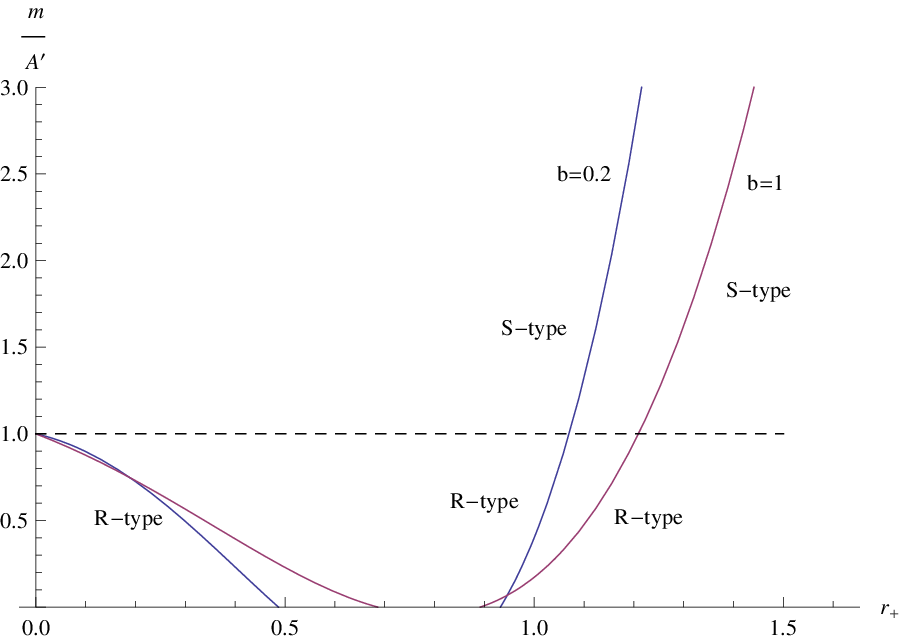}}%
\hfill%
\subfigure[$k=0$]{
\label{fig:subfig:c} %% label for second subfigure
\includegraphics[width=0.3\textwidth]{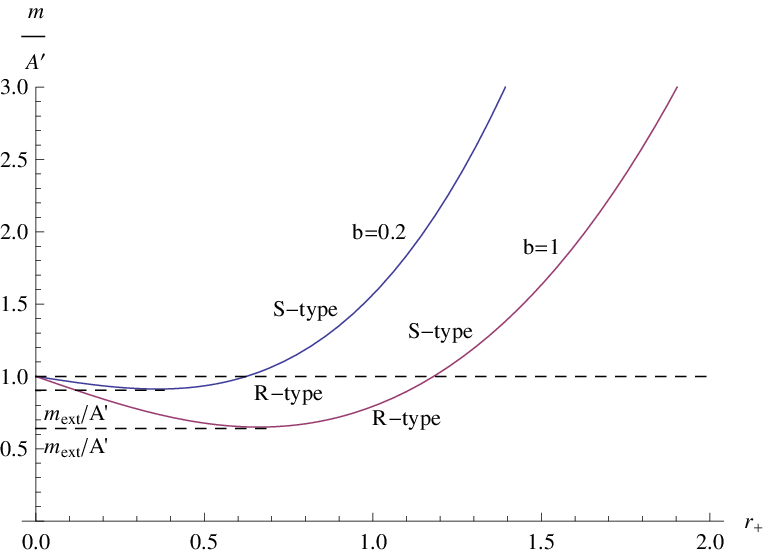}}
\caption{Solutions of black hole horizons in
five-dimensions as functions of $m/A'$. We take
$Q=1$, $l=1$. }
\label{fig:subfig} %% label for entire figure
\end{figure}

%%%%%%%%%%%%%%%%%%%%%%%%%%%%%%%%%%%%%%%%%%%%%%%%%%%%%%%%%%%%%%%%%%%%%%%%%%%
\begin{figure}[h]
\centering
\subfigure[$k=1$]{
\label{fig:subfig:a} %% label for first subfigure
\includegraphics[width=0.3\textwidth]{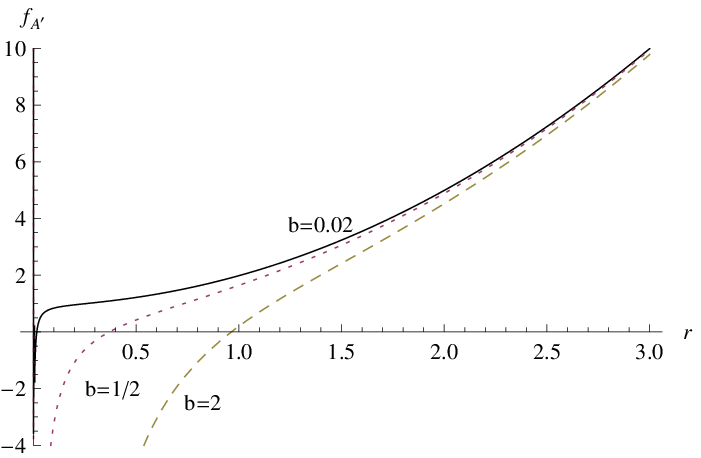}}%
\hfill%
\subfigure[$k=-1$]{
\label{fig:subfig:b} %% label for first subfigure
\includegraphics[width=0.3\textwidth]{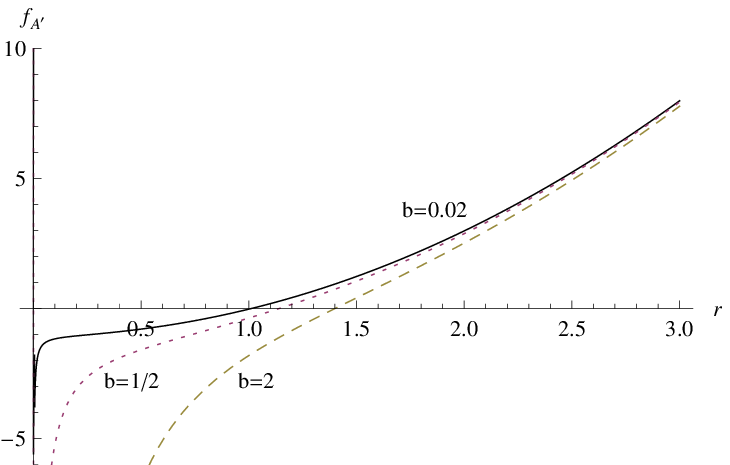}}%
\hfill%
\subfigure[$k=0$]{
\label{fig:subfig:c} %% label for second subfigure
\includegraphics[width=0.3\textwidth]{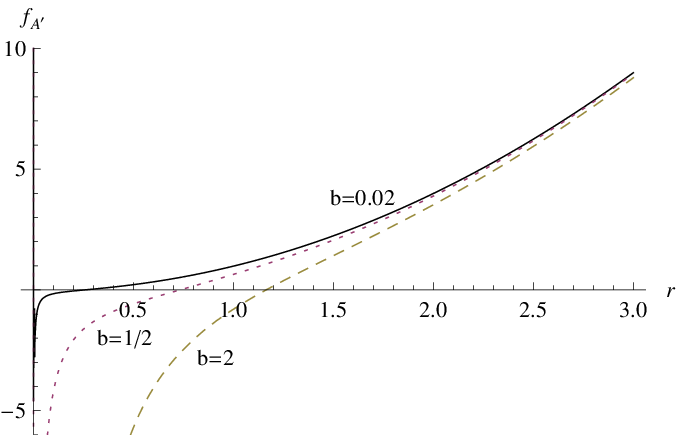}}
\caption{ Marginal case in five-dimensions. We
fix $Q=1$, $l=1$.}
\label{fig:subfig} %% label for entire figure
\end{figure}

%%%%%%%%%%%%%%%%%%%%%%%%%%%%%%%%%%%%%%%%%%%%%%%%%%%%%%%%%%%%%%%%%%%%%%%%%%%
\begin{figure}[htb]
\centering
\subfigure[$k=1$]{\label{fig:a} %% label for second subfigure
\includegraphics[width=0.3\textwidth]{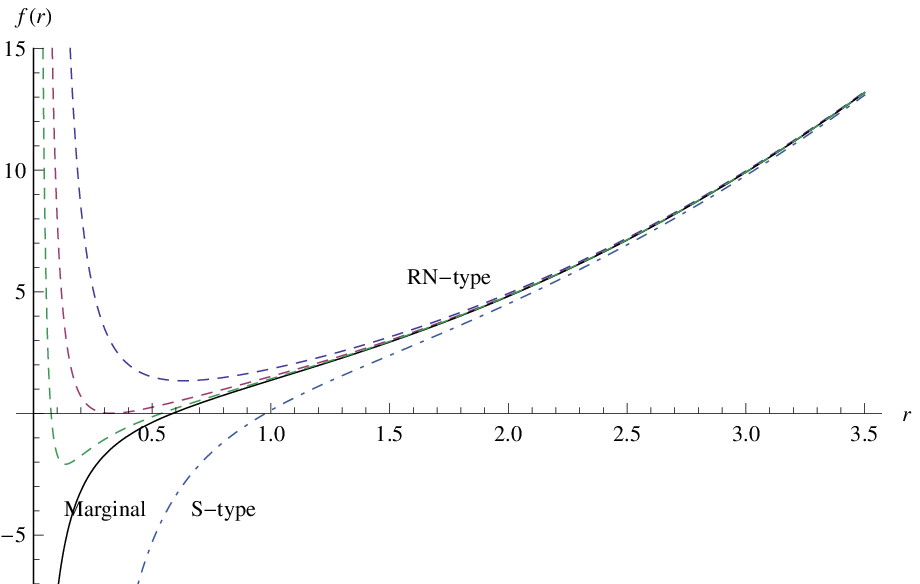}}%
\hfill%
\subfigure[$k=-1$]{\label{fig:b} %% label for second subfigure
\includegraphics[width=0.3\textwidth]{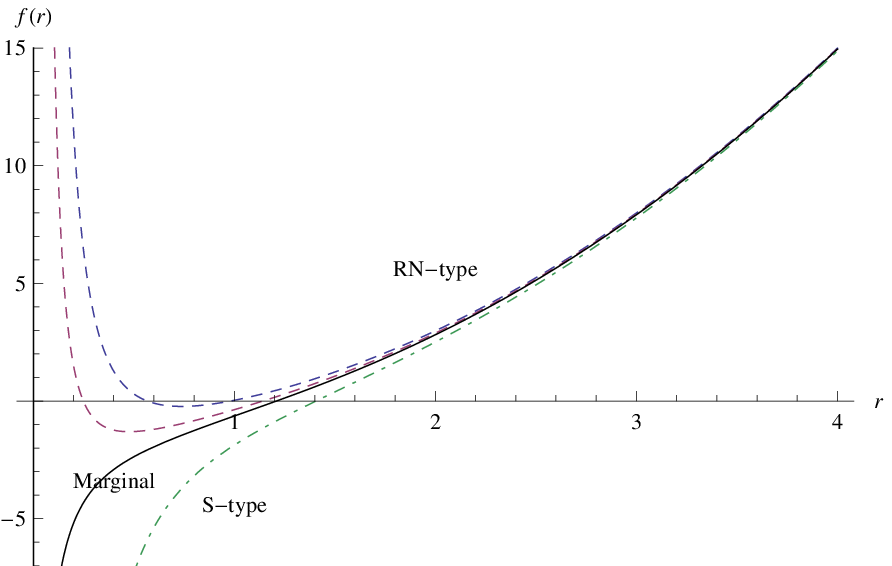}}%
\hfill%
\subfigure[$k=0$]{\label{fig:c} %% label for second subfigure
\includegraphics[width=0.3\textwidth]{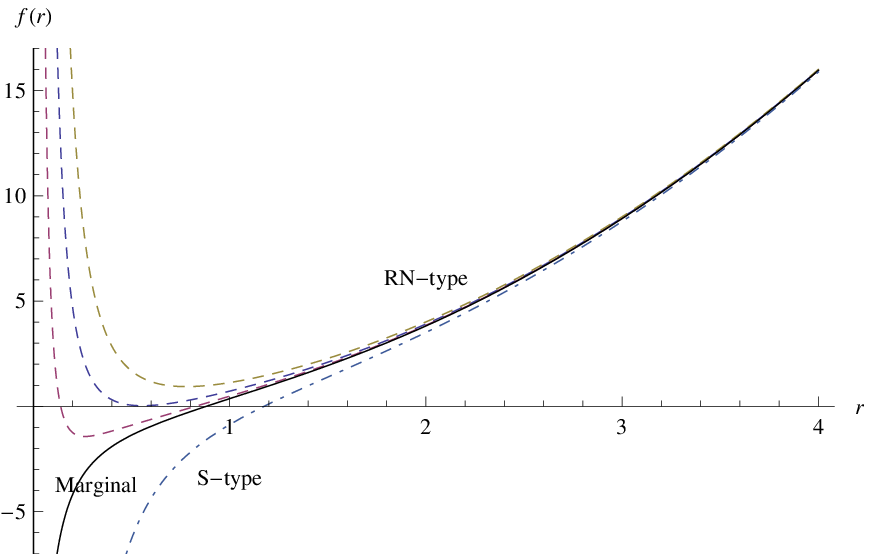}}%
\caption{Types of the
five-dimensional Born-Infeld AdS black holes with
different topologies. Here the ``marginal" case $(m=A')$
belongs to the S-type. For the RN-type black hole $(m<A')$ with $k=1$ and 0, the three
dashed lines from top to below correspond to the naked singularity taking $m<m_{ext}<A'$,
extremal BH $m=m_{ext}<A'$ and two horizon solutions
$m_{ext}<m<A'$ respectively. However, the RN-type of Born-Infeld
AdS black hole with $k=-1$ always possesses two
horizons. We take $Q=1$, $b=1$.}
\label{fig:subfig} %% label for entire figure
\end{figure}

\section{Phase transitions of $D$-dimensional Born-Infeld AdS black holes}
\label{3s}

\subsection{Phase transitions }

In the geometric units $G_N=\hbar=c=k=1$, we interpret the cosmological constant
$\Lambda=-\frac{(D-1)(D-2)}{2l^2}$ as a positive thermodynamic pressure
\begin{eqnarray}
P=-\frac{1}{8\pi}\Lambda=\frac{(D-1)(D-2)}{16\pi l^2}.\label{eq:19a}
\end{eqnarray}
The black hole mass $M$ can be considered as the enthalpy rather than the internal energy of the
gravitational system \cite{Kastor:2009wy}. In the Born-Infeld case, $M$ should be the function of
entropy, pressure, charge, and Born-Infeld coupling coefficient \cite{Gunasekaran:2012dq}.
Moreover, those thermodynamic quantities satisfy the following differential form
\begin{eqnarray}
dM=TdS+\Phi dQ+VdP+\mathfrak{B}db,\label{eq:20a}
\end{eqnarray}
where the thermodynamic volume $V$ conjugate to $P$ equals to $\frac{\Sigma_k r_+^{D-1}}{D-1}$.
$\mathfrak{B}$ is a quantity conjugate to $b$ and is called the `Born-Infeld vacuum polarization'
\begin{eqnarray}
\mathfrak{B}&=&\frac{\Sigma_k b r_+^{D-1}}{2(D-1)\pi}\left(1-\sqrt{1+\frac{16\pi^2 Q^2}{b^2\Sigma_k^2 r_+^{2D-4}}}\right)\nonumber\\
&&+\frac{4Q^2\pi}{(D-1)b\Sigma_k r_+^{D-3}}{_2}F_1[\frac{D-3}{2D-4},\frac{1}{2},\frac{3D-7}{2D-4},
-\frac{16\pi^2 Q^2}{b^2\Sigma_k^2 r_+^{2D-4}}].\label{eq:21a}
\end{eqnarray}
For $D=4$, Eq.~(\ref{eq:21a}) reduces to \cite{Gunasekaran:2012dq}
\begin{eqnarray}
\mathfrak{B}=\frac{2b r_+^{3}}{3}\left(1-\sqrt{1+\frac{Q^2}{b^2r_+^{4}}}\right)
+\frac{Q^2}{3b r_+}{_2}F_1[\frac{1}{4},\frac{1}{2},\frac{5}{4},-\frac{Q^2}{b^2r_+^{4}}].\label{eq:22a}
\end{eqnarray}
By scaling argument, we can obtain the generalized Smarr relation for the Born-Infeld
AdS black holes in the extended phase space
\begin{eqnarray}
M=\frac{D-2}{D-3}TS+\Phi Q-\frac{2}{D-3}VP-\frac{1}{D-3}\mathfrak{B}b.\label{eq:23a}
\end{eqnarray}
In four dimensional spacetime, this expression reduces to that described in \cite{Gunasekaran:2012dq}.

Using Eqs.(\ref{eq:9a})(\ref{eq:19a}), the equation of state $P(V,T)$ can be obtained
\begin{eqnarray}
P=\frac{(D-2)T}{4r_+}-\frac{(D-2)(D-3)k}{16\pi r_+^2}
-\frac{b^2}{4\pi}\left(1-\sqrt{1+\frac{16\pi^2 Q^2}{b^2\Sigma_k^2 r_+^{2D-4}}}\right).\label{eq:24a}
\end{eqnarray}
To compare with the Van der Waals fluid equation in $D$-dimensions, we can translate the
``geometric" equation of state to physical one by identifying the specific volume $v$ of the fluid
with the horizon radius of the black hole $r_+$ as $v=\frac{4r_+}{D-2}$ \cite{Kubiznak:2012wp, Gunasekaran:2012dq}.

We know that the critical point occurs when $P=P(v)$ has an inflection point
\begin{eqnarray}
\frac{\partial P}{\partial r_+}\Big|_{T=T_c, r_+=r_c}=0, \quad
\frac{\partial^2 P}{\partial r_+^2}\Big|_{T=T_c, r_+=r_c}=0.\label{eq:25a}
\end{eqnarray}
Then we can obtain the critical temperature
\begin{eqnarray}
T_c=\frac{(D-3)k}{2\pi r_c}-\frac{16\pi Q^2}{\Sigma_k^2 r_c^{2D-5}}\left(1
+\frac{16\pi^2 Q^2}{b^2\Sigma_k^2 r_c^{2D-4}}\right)^{-1/2} \label{eq:26a}
\end{eqnarray}
and the equation for the critical horizon radius
$r_c$ (specific volume $v_c=\frac{4r_c}{D-2}$)
\begin{eqnarray}
F(r_c)&=&k-\frac{32(2D-5)\pi^2Q^2}{(D-3)\Sigma_k^2r_c^{2D-6}}\left(1
+\frac{16\pi^2 Q^2}{b^2\Sigma_k^2 r_c^{2D-4}}\right)^{-1/2}\nonumber\\
&&+\frac{512(D-2)\pi^4 Q^4}{(D-3) b^2\Sigma_k^4 r_c^{4D-10}}\left(1
+\frac{16\pi^2 Q^2}{b^2\Sigma_k^2 r_c^{2D-4}}\right)^{-3/2}=0,\label{eq:27a}
\end{eqnarray}
where $r_c$ denotes the critical value of $r_+$.

For different topological spacetimes, we can
investigate the phase structure and $P-V$
criticality in the extended phase space.
Obviously the critical temperature $T_c$
Eq.~(\ref{eq:26a}) is negative for hyperbolic and
flat black hole horizons, so that no phase transition
happens there. Thus we only need to explore the
phase transition and $P-V$ criticality in the
spherical Born-Infeld AdS black hole. The behaviors of $F(r_c)$ for
different values of $b$ in different spacetime
dimensions can be seen in Fig.9.

%%%%%%%%%%%%%%%%%%%%%%%%%%%%%%%%%%%%%%%%%%%%%%%%%%%%%%%%%%%%%%%%%%%%%%%%%%%
\begin{figure}[htb]
\centering
\subfigure[$D=4$]{
\label{fig:subfig:a} %% label for first subfigure
\includegraphics[width=0.3\textwidth]{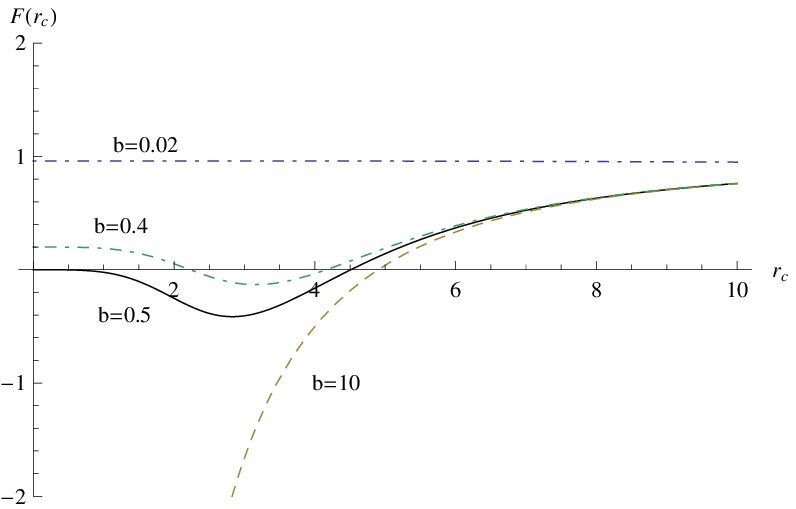}}%
\hfill%
\subfigure[$D=5$]{
\label{fig:subfig:b} %% label for first subfigure
\includegraphics[width=0.3\textwidth]{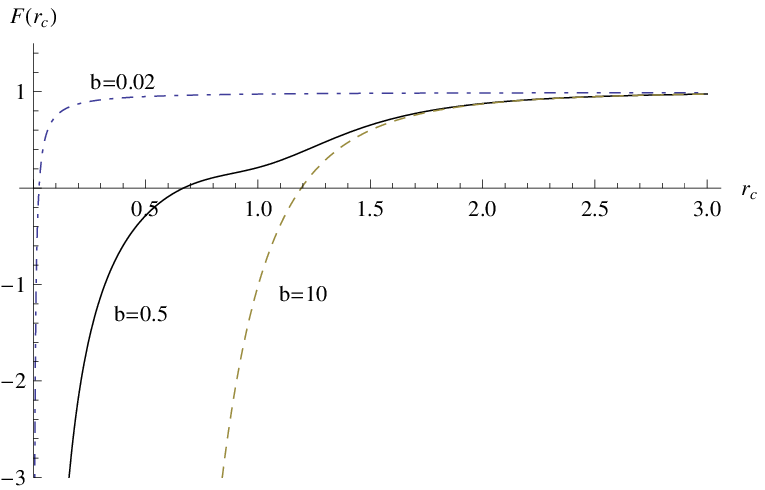}}%
\hfill%
\subfigure[$D=6$]{
\label{fig:subfig:c} %% label for second subfigure
\includegraphics[width=0.3\textwidth]{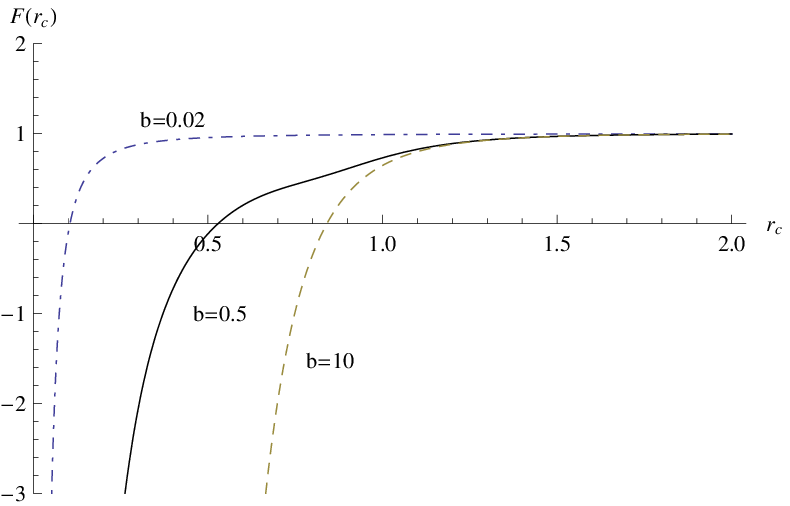}}
\caption{ The behaviors of $F(r_c)$ for various
values of $b$ in different spacetime dimensions.
We fix $Q=1$, $l=1$ and $k=1$.}
\label{fig:subfig} %% label for entire figure
\end{figure}

When $D=4$, the critical point was discussed in details in \cite{Gunasekaran:2012dq}. Rewriting
Eq.~(\ref{eq:27a}) with $v_c=2r_c$, we have
\begin{eqnarray}
x^{3}-\frac{3b^2}{32Q^2}x+\frac{b^2}{256Q^4}=0, \quad x=\left(v_c^{4}+\frac{16Q^2}{b^2}\right)^{-1/2}, \label{eq:28a}
\end{eqnarray}
where it has one or more positive real roots for $r_c$ and displays different phase transitions of
four dimensional Born-Infeld AdS black hole. For four dimensions, these results can also
be obtained by analyzing the function $\frac{\partial F(r_c)}{\partial r_c}$
\begin{eqnarray}
\frac{\partial F(r_c)}{\partial r_c}=\frac{12Q^2}{r_c^{3}}\left(1+\frac{Q^2}{b^2r_c^{4}}\right)^{-5/2}
 \left(1-\frac{Q^2}{b^2 r_c^{4}}\right),\label{eq:29a}
\end{eqnarray}
which can take positive, zero or negative values depending on the values of $b$ and $Q$. We
summarize real roots of Eq.~(\ref{eq:29a}) and the critical points in the TABLE.~\Rmnum{2} \cite{Gunasekaran:2012dq}.

\begin{table}[h]
\begin{tabular}{|c||c||c||c||c|}
  \hline
  \multicolumn{5}{|c|}{Born-Infeld AdS black holes} \\ \hline
  Coupling coefficient & $b>\frac{1}{2Q} $ & $b<\frac{1}{\sqrt{8}Q}$ & $\frac{1}{\sqrt{8}Q}< b<\frac{\sqrt{3+2\sqrt{3}}}{6Q}$
  & $\frac{\sqrt{3+2\sqrt{3}}}{6Q}< b<\frac{1}{2Q}$   \\ \hline
  Number of real root $x$ & one & one & three & three   \\ \hline
  Number of critical points & one & none & two & one \\ \hline
  types of BI-AdS BH & RN & S & S & S \\ \hline
\end{tabular}
\caption{The behaviors of critical points for different values of $b$ in four dimensions by setting $Q=1$.}
\end{table}

When the spacetime dimension $D \geq 5$, it is difficult to give analytical roots of $F(r_c)=0$.
The number of real roots of Eq.~(\ref{eq:27a}) can be decided by examining
\begin{eqnarray}
\frac{\partial F(r_c)}{\partial r_c}&=&\frac{64(2D-5)\pi^2Q^2}{\Sigma_k^2 r_c^{2D-5}}\left(1
+\frac{16\pi^2 Q^2}{b^2\Sigma_k^2 r_c^{2D-4}}\right)^{-5/2}\nonumber\\
&&\times\left[1+\frac{8(D-6)\pi^2Q^2}{(D-3)b^2\Sigma_k^2
r_c^{2D-4}}+\frac{128(D-4)\pi^4Q^4}{(2D-5)b^4\Sigma_k^4 r_c^{4D-8}}\right].\label{eq:30a}
\end{eqnarray}
Taking $D=5$, we have
\begin{eqnarray}
&&1+\frac{8(D-6)\pi^2Q^2}{(D-3)b^2\Sigma_k^2 r_c^{2D-4}}+\frac{128(D-4)\pi^4Q^4}{(2D-5)b^4\Sigma_k^4 r_c^{4D-8}}\nonumber\\
&&=\left(1+\frac{4(D-6)\pi^2Q^2}{(D-3)b^2\Sigma_k^2r_c^{2D-4}}\right)^2
+\frac{48(2D-9)(D-2)^2\pi^4Q^4}{(2D-5)(D-3)^2b^4\Sigma_k^4 r_c^{4D-8}}>0,\label{eq:31a}
\end{eqnarray}
so that $\frac{\partial F(r_c)}{\partial r_c}$ is
always positive for $D=5$ and there is only one
real root for $F(r_c)=0$. Thus the equation of
state Eq.~(\ref{eq:24a}) allows only one critical
point for each value of $b$ when $D=5$. This
property keeps for $D\geq 6$, since the function
$\frac{\partial F(r_c)}{\partial r_c}$ is always
positive, so that only one critical point exists
for $D\geq 6$. Compared with
the four-dimensional case, the phase structure
for the higher dimensional case is much simpler.

We can numerically solve Eq.~(\ref{eq:27a}). We list the phase space parameters at critical
points for different values of $b$ when spacetime dimension $D=5$ and $6$ in TABLE.~\Rmnum{3}. We
plot the $P-V$ diagrams for $D=5$ and 6 in Fig.~10. The two upper dashed lines correspond to
the ``ideal gas" phase behavior when $T>T_c$. For $T<T_c$, there is a small-large black hole phase
transition in the system. Different from the four-dimensional Born-Infeld AdS black hole case,
we have not observed the reentrant phase transition when the spacetime dimension is higher.

\begin{table}[h]
\begin{tabular}{|c||c||c||c||c||c||c||c||c||c||c|}
  \hline
   parameters & \multicolumn{5}{|c|}{D=5} & \multicolumn{5}{|c|}{D=6}\\ \hline
  $b $ & 0.02 & 0.1& 1 & 10 & 100 & 0.02 & 0.1& 1 & 10 & 100 \\ \hline
  $v_c$ &0.0340 & 0.1698& 1.5196 & 1.5902 & 1.5908 &0.1050 & 0.2350& 0.7813 & 0.8409& 0.8413 \\ \hline
  $T_c$ & 6.2500 & 1.2500&  0.2177 & 0.2135& 0.2134 & 3.0313 & 1.3560& 0.5017 & 0.4866& 0.4864 \\ \hline
  $P_c$ & 61.3592& 2.4536 & 0.0584 & 0.0565 & 0.0559 & 10.8253 & 2.1643& 0.2704  & 0.2531& 0.2530 \\ \hline
  $\frac{P_c v_c}{T_c}$ & 0.3333& 0.3333& 0.4076 & 0.4147 & 0.4167 & 0.3750& 0.3750& 0.4212 & 0.4374 & 0.4375\\ \hline
\end{tabular}
\caption{Phase space parameters at critical points for different values of $b$ when spacetime
dimension $D=5$ and $6$. We fix $Q=1$. As $b\rightarrow \infty$, the value of $\frac{P_c
v_c}{T_c}$ approaches the critical value of 5D and 6D Reissner-N\"{o}rdstrom black holes \cite{Gunasekaran:2012dq}.}
\end{table}

%%%%%%%%%%%%%%%%%%%%%%%%%%%%%%%%%%%%%%%%%%%%%%%%%%%%%%%%%%%%%%%%%%%%%%%%%%%
\begin{figure}[htb]
\centering
\subfigure[$D=5$ and $b=10$]{\label{fig:a} %% label for second subfigure
\includegraphics[width=0.35\textwidth]{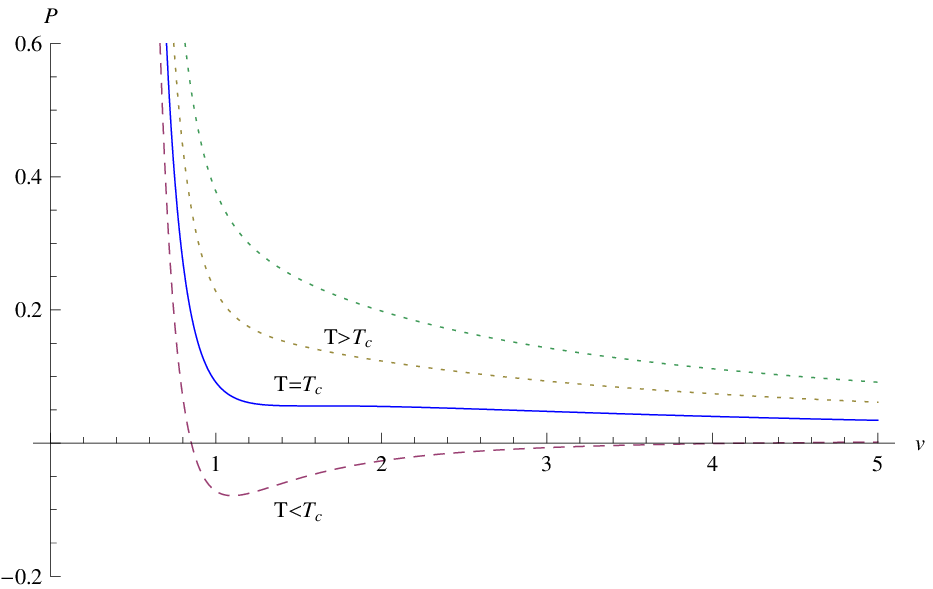}}%
\hfill%
\subfigure[$D=5$ and $b=0.1$]{\label{fig:b} %% label for second subfigure
\includegraphics[width=0.35\textwidth]{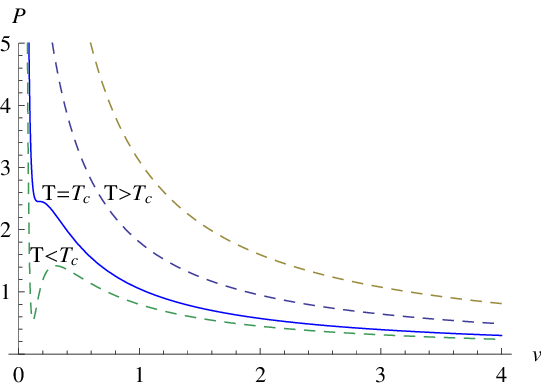}}%
\hfill%
\subfigure[$D=6$ and $b=10$]{\label{fig:c} %% label for second subfigure
\includegraphics[width=0.35\textwidth]{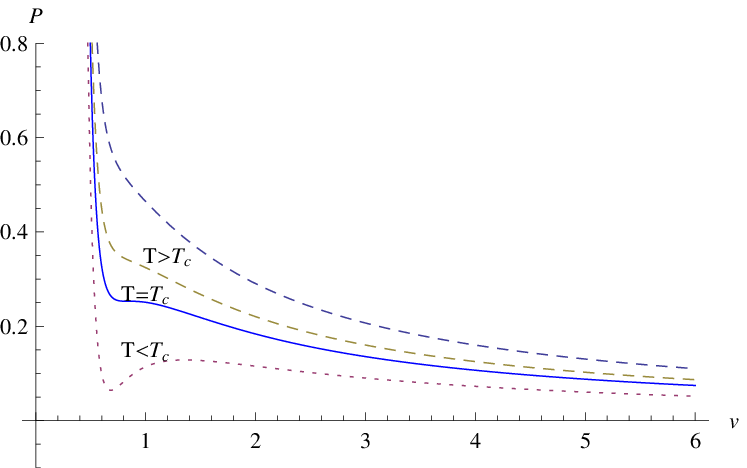}}%
\hfill%
\subfigure[$D=6$ and $b=0.1$]{\label{fig:d} %% label for second subfigure
\includegraphics[width=0.35\textwidth]{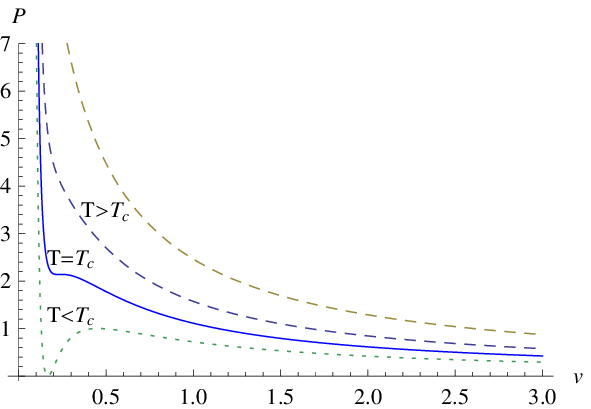}}
\caption{ The P-V diagram of Born-Infeld AdS
black holes when the spacetime dimensions
$D=5,6$. We fix $Q=1$. The two upper dashed lines
correspond to the ``idea gas" phase behavior for
$T>T_c$. The critical temperature case is denoted
by the solid line. The lines below are with
temperatures smaller than the critical
temperature.}
\label{fig:subfig} %% label for entire figure
\end{figure}

To get more information about the phase
transition, we can examine the free energy. For
the canonical ensemble, the Gibbs free energy is
$G=M-TS$, which reads
\begin{eqnarray}
G(T,P)&=&\frac{\Sigma_k}{16\pi}\left[kr_+^{D-3}-\frac{16\pi P r_+^{D-1}}{(D-1)(D-2)}-\frac{4b^2r_+^{D-1}}{(D-1)(D-2)}\left(1
-\sqrt{1+\frac{16\pi^2 Q^2}{b^2\Sigma_k^2 r_+^{2D-4}}}\right)\right.\nonumber\\
&&\left.+\frac{64(D-2)^2\pi^2Q^2}{(D-1)(D-3)\Sigma_k^2r_+^{D-3}}{_2}F_1[\frac{D-3}{2D-4},\frac{1}{2},\frac{3D-7}{2D-4},
-\frac{16\pi^2 Q^2}{b^2\Sigma_k^2 r_+^{2D-4}}]\right].\label{eq:32a}
\end{eqnarray}
Here $r_+$ is understood as the function of
pressure and temperature, $r_+=r_+(P,T)$, via
equation of state Eq.~(\ref{eq:24a}). We plot the
change of the free energy $G$ with $T$ for fixed
$Q$ and different spacetime dimensions in
Fig.~11, which reveals the existence of ``swallow
tail" behavior of the free energy $G$. The
presence of the characteristic ``swallow tail"
behavior means that the small-large black hole
phase transition occurs in the system is of the
first order. The coexistence line in the
$(P,T)$-plane describes that two phases share the
same Gibbs free energy and temperature during the
phase transition, which is plotted in Fig.~12.
Note that this coexistence line looks very
similar to the van de Waals liquid-gas system.

%%%%%%%%%%%%%%%%%%%%%%%%%%%%%%%%%%%%%%%%%%%%%%%%%%%%%%%%%%%%%%%%%%%%%%%%%%%
\begin{figure}[htb]
\centering
\subfigure[$D=5$ and $b=10$]{\label{fig:subfig:a} %% label for second subfigure
\includegraphics[width=0.35\textwidth]{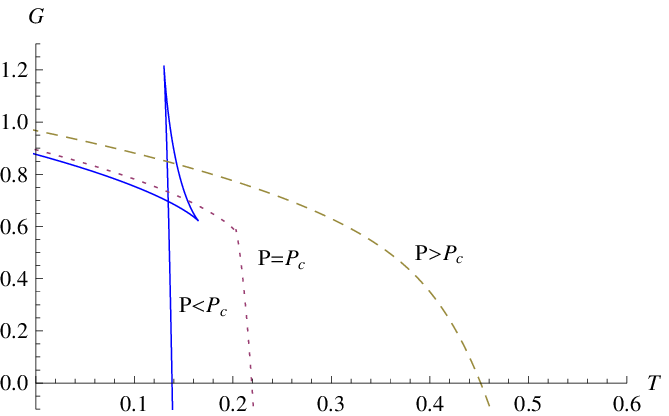}}%
\hfill%
\subfigure[$D=5$ and $b=0.1$]{\label{fig:subfig:b} %% label for second subfigure
\includegraphics[width=0.35\textwidth]{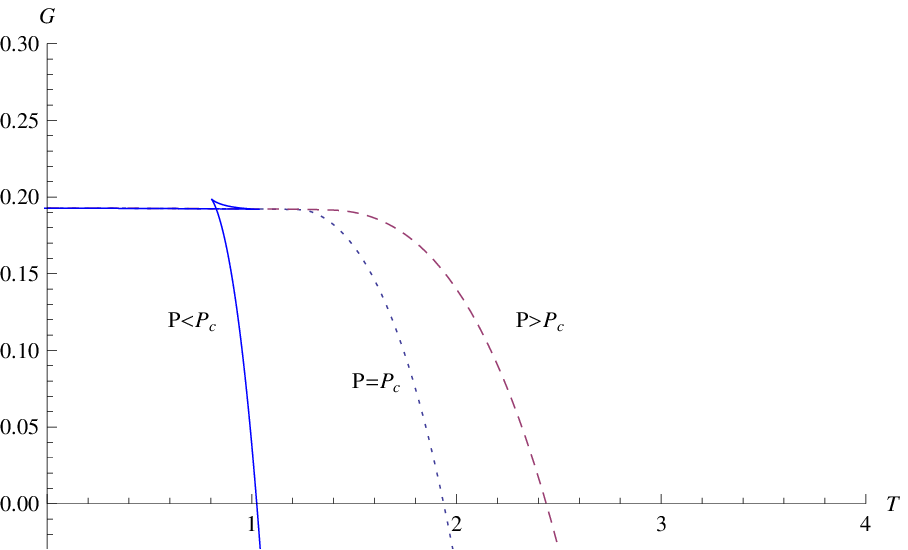}}%
\hfill%
\subfigure[$D=6$ and $b=10$]{\label{fig:subfig:c} %% label for second subfigure
\includegraphics[width=0.35\textwidth]{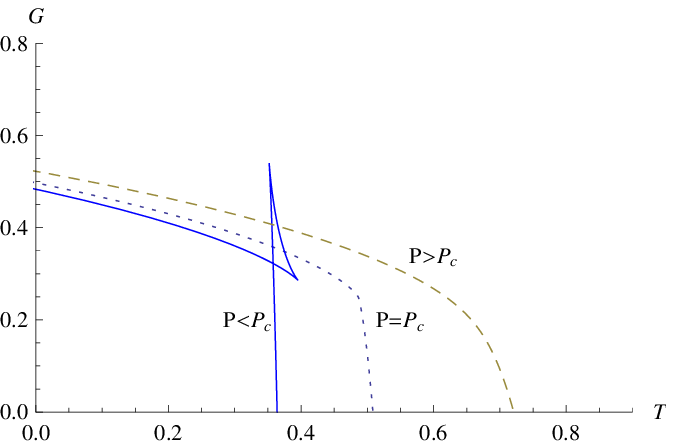}}%
\hfill%
\subfigure[$D=6$ and $b=0.1$]{\label{fig:subfig:d} %% label for second subfigure
\includegraphics[width=0.35\textwidth]{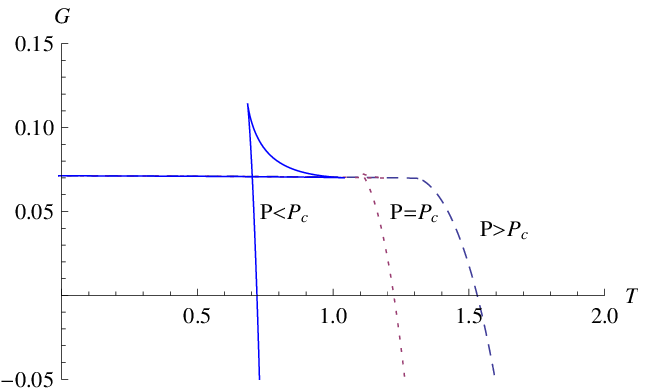}}
\caption{ The Gibbs free energy as a function of
temperature different spacetime dimensions
$D=5,6$. We fix $Q=1$.}
\label{fig:subfig} %% label for entire figure
\end{figure}

%%%%%%%%%%%%%%%%%%%%%%%%%%%%%%%%%%%%%%%%%%%%%%%%%%%%%%%%%%%%%%%%%%%%%%%%%%%
\begin{figure}[htb]
\centering
\subfigure[$b=0.1$]{\label{fig:subfig:a} %% label for second subfigure
\includegraphics[width=0.35\textwidth]{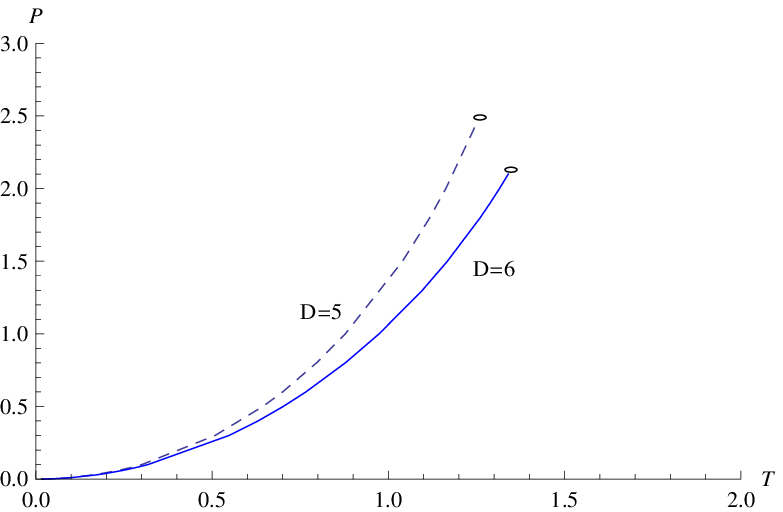}}%
\hfill%
\subfigure[$b=10$]{\label{fig:subfig:b} %% label for second subfigure
\includegraphics[width=0.35\textwidth]{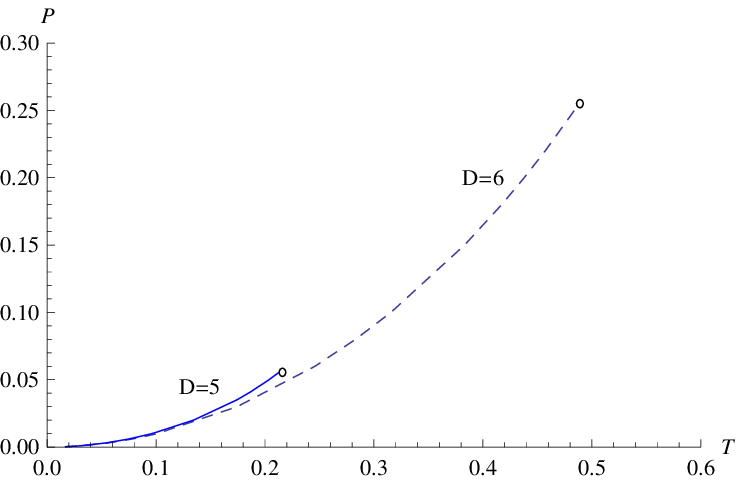}}%
\caption{ The coexistence line of small-large
Born-Infeld AdS black hole phase transition for
different dimensions in $(P,T)$-plane. The
critical point is shown by a small circle at the
end of the coexistence line.}
\label{fig:subfig} %% label for entire figure
\end{figure}

Now we turn to compute the critical exponents characterizing the behavior of physical quantities in the
vicinity of the critical point in the Born-Infeld AdS black hole system. Although the equation of state
$P(V,T)$ Eq.~(\ref{eq:24a}) is dimensional dependence, we will show this does not affect the behavior of
the critical exponents. Near the critical point, the critical behavior
of van de Waals liquid-gas system can be characterized by the following critical exponents \cite{Kubiznak:2012wp},

\begin{eqnarray}
&&C_v=T\frac{\partial S}{\partial T}\Big|_v\propto \left(-\frac{T-T_c}{T_c}\right)^{-\alpha},\nonumber\\
&&\eta=\frac{v_s-v_l}{v_c}\propto \left(-\frac{T-T_c}{T_c}\right)^{\beta},\nonumber\\
&&\kappa_T=-\frac{1}{v}\frac{\partial v}{\partial P}\Big|_T\propto \left(-\frac{T-T_c}{T_c}\right)^{-\gamma},\nonumber\\
&& P-P_c \propto (v-v_c)^{\delta}.\label{eq:33a}
\end{eqnarray}

In order to compute the critical exponent $\alpha$, the entropy $S$ of Born-Infeld AdS black hole Eq.~(\ref{eq:10a}) can be
rewritten as $\frac{\Sigma_k^{\frac{1}{D-1}}}{4}\left((D-1)V\right)^{\frac{D-2}{D-1}}$. Obviously this entropy $S$ is independent
of $T$ and hence we have the critical exponent $\alpha=0$. To obtain other exponents, define
\begin{eqnarray}
p=\frac{P}{P_c}, \quad \nu=\frac{v}{v_c}, \quad \tau=\frac{T}{T_c} \label{eq:34a}
\end{eqnarray}
and introduce the expansion parameters
\begin{eqnarray}
\tau=t+1, \quad \nu=\omega+1,\label{eq:35a}
\end{eqnarray}
then the expansion for this equation of state near the critical point is given by
\begin{eqnarray}
p=1+a_{10}t+a_{11}t\omega+a_{03}\omega^3+\mathcal{O}(t\epsilon^2,\epsilon^4).\label{eq:36a}
\end{eqnarray}
During the phase transition, the pressure remains constant
\begin{eqnarray}
&&p=1+a_{10}t+a_{11}t\omega_s+a_{03}\omega_s^3=1+a_{10}t+a_{11}t\omega_l+a_{03}\omega_l^3, \nonumber\\
\Rightarrow && a_{11}t\left(\omega_s-\omega_l\right)+a_{03}\left(\omega_s^3-\omega_l^3\right)=0,\label{eq:37a}
\end{eqnarray}
where $\omega_s$ and $\omega_l$ denote the `volume' of small and large black holes.
Using Maxwell's area law, we can obtain
\begin{eqnarray}
\int^{\omega_s}_{\omega_l}\omega\frac{dp}{d\omega}d\omega=0 \Rightarrow a_{11}\left(\omega_s^2-\omega_l^2\right)
+\frac{3}{2}a_{03}\left(\omega_s^4-\omega_l^4\right)=0.\label{eq:38a}
\end{eqnarray}
The nontrivial solution of Eqs.~(\ref{eq:37a})(\ref{eq:38a}) appears only when $a_{11}a_{03}t<0$, reads as
\begin{eqnarray}
\omega_s=\frac{\sqrt{-a_{11}a_{03}t}}{3|a_{03}|}, \quad \omega_l=-\frac{\sqrt{-a_{11}a_{03}t}}{3|a_{03}|}.\label{eq:39a}
\end{eqnarray}
In general, it seems impossible to give an analytical expression for $a_{11}$ and $a_{03}$.
The numerically results of $a_{10}$, $a_{11}$ and $a_{03}$ for different $b$ are shown in TABLE.~\Rmnum{4}.
Therefore we have
\begin{eqnarray}
\eta=\omega_s-\omega_l=2\omega_s\propto \sqrt{-t}\Rightarrow \beta=1/2.\label{eq:40a}
\end{eqnarray}

\begin{table}[h]
\begin{tabular}{|c||c||c||c||c||c||c||c||c||c||c|}
  \hline
   parameters & \multicolumn{5}{|c|}{D=5} & \multicolumn{5}{|c|}{D=6}\\ \hline
  $b $ & 0.02 & 0.1& 1 & 10 & 100 & 0.02 & 0.1& 1 & 10 & 100 \\ \hline
  $a_{10}$ &2.9994 & 3.0003& 2.4534 & 2.4004 & 2.4004 &2.6667 & 2.6667& 2.3742 & 2.2862 & 2.2857 \\ \hline
  $a_{11}$ & -2.9994 & -3.0003& -2.4534 & -2.4004 & -2.4004 & -2.6667 & -2.6667& -2.3742 & -2.2862 & -2.2857 \\ \hline
  $a_{03}$ &-0.9911& -0.9885&-1.5791 & -1.9967 & -2.0012 & -1.3341 & -1.3240& -1.5787 & -2.6550 & -2.6672 \\ \hline
\end{tabular}
\caption{The values of $a_{10}$, $a_{11}$ and $a_{03}$ with different values of $b$ for $D=5,6$ and $Q=1$.}
\end{table}

The isothermal compressibility can be computed as
\begin{eqnarray}
\kappa_T=-\frac{1}{v}\frac{\partial v}{\partial P}\Big|_{v_c}\propto
-\frac{1}{\frac{\partial p}{\partial \omega}}\Big|_{\omega=0}=-\frac{1}{a_{11}t},\label{eq:41a}
\end{eqnarray}
which indicates that the critical exponent $\gamma=1$. In addition, we obtain the shape of the critical
isotherm $t=0$,
\begin{eqnarray}
p-1=a_{03}\omega^3\Rightarrow \delta=3.\label{eq:42a}
\end{eqnarray}
It is easy to check that these critical exponents satisfy the following thermodynamic scaling laws
\begin{eqnarray}
&&\alpha+2\beta+\gamma=2, \quad \alpha+\beta(1+\delta)=2,\nonumber\\
&&\gamma(1+\delta)=(2-\alpha)(\delta-1), \quad \gamma=\beta(\delta-1).\label{eq:43a}
\end{eqnarray}
Therefore these thermodynamic exponents for the higher dimensional Born-Infeld AdS black holes
coincide with those of the Van de Waals fluid \cite{Kubiznak:2012wp}, and also take the same values
in the (higher dimensional) Maxwell case \cite{Gunasekaran:2012dq}.

\subsection{Phase transition of three dimensional Born-Infeld AdS black hole}

We extend our investigate to the three
dimensional Born-Infeld black holes. The metric
is given by \cite{Cataldo:1999wr, Myung:2008kd}
\begin{eqnarray}
ds^2&=&-f(r)dt^2+f(r)^{-1}dr^2+r^2 d\theta^2,\nonumber\\
F&=&E(r)dt\wedge dr, \quad E(r)=\frac{Q}{\sqrt{r^2+Q^2/b^2}}\label{eq:44a}
\end{eqnarray}
with
\begin{eqnarray}
f(r)&=&-m+\frac{r^2}{l^2}+2b^2 r\left(r-\sqrt{r^2+Q^2/b^2}\right)-2Q^2\ln\left(r+\sqrt{r^2+Q^2/b^2}\right)\nonumber\\
&&+2Q^2\ln\left(l+\sqrt{l^2+Q^2/b^2}\right)-2b^2l\left[l-\sqrt{l^2+Q^2/b^2}\right].\label{eq:45a}
\end{eqnarray}
The temperature of black hole is
$T=\frac{2r_+}{l^2}+4b^2\left(r_{+}-\sqrt{r_+^2+Q^2/b^2}\right)$.
Taking $\Lambda=-1/l^2=-8\pi P$, we can have the
equation of state
\begin{eqnarray}
P=\frac{T}{16\pi r_+}-\frac{b^2}{4\pi}\left(1-\sqrt{1+\frac{Q^2}{b^2r_+^2}}\right). \label{eq:46a}
\end{eqnarray}
According to Eq.~(\ref{eq:25a}), one can find that such an equation does not admit any
inflection point. Hence the three dimensional Born-Infeld black hole does not exhibit any
critical behavior. This situation agrees with the finding in three-dimensional BTZ black hole
\cite{Gunasekaran:2012dq} and three-dimensional black hole with scalar field \cite{Belhaj:2013ioa}.

\section{closing remarks}
\label{4s}

In this paper we have studied the thermodynamic
behaviors of $D$-dimensional Born-Infeld AdS
black holes in an extended phase space by
treating the cosmological constant and the
maximal field strength $b$ and their conjugate
quantities as thermodynamic variables,
respectively. We have written out the equations
of state and examined the phase structures by
using the standard thermodynamic techniques. We
have shown that systems with dimensions higher
than four admit a first order small-large black
hole phase transition which resembles the
liquid-gas phase transition in fluids. We have
not found the reentrant phase transition, which
was observed in four-dimensional Born-Infeld AdS
black holes, for higher-dimensional spacetimes.
In addition, in the three dimensional Born-Infeld
AdS black hole, we have shown that there does not
exist any critical phenomena. This agrees with
the findings in other three-dimensional AdS black
holes.

{\bf Acknowledgments}

This work was supported by the National Natural Science Foundation of China.

\end{document}